%% file: AMS_DYSM.tex
\documentclass[11 pt, a4paper]{article}

\newcommand{\btheta}{{\boldsymbol{\tilde{\vartheta}}}}

\input{pack.tex}

\usepackage{blindtext}
\usepackage[pagewise]{lineno}

\usepackage[round]{natbib}
\usepackage{tikz}
\usetikzlibrary{fit,positioning,calc}

\newcommand*\dif{\mathop{}\!\mathrm{d}}
\newcommand{\name}[0]{\textsc{dysm}} 
\newcommand{\simI}{\overset{\mbox{\tiny iid}}{\sim}} 

\title{Dynamic modeling of mortality via mixtures of skewed distribution functions}
\author[1]{Emanuele Aliverti}
\author[2]{Stefano Mazzuco}
\author[2,3]{Bruno Scarpa}
\affil[1]{\small\emph{Department of Economics, University Ca' Foscari Venezia}}
\affil[2]{\small\emph{Department of Statistical Sciences, University of Padova}}
\affil[3]{\small\emph{Department of Mathematics ``Tullio Levi-Civita'', University of Padova}}

\date{}

\begin{document}
\maketitle
\begin{abstract}
	There has been growing interest on forecasting mortality.
In this article, we propose a novel dynamic Bayesian approach for modeling and forecasting the age-at-death distribution, focusing on a three-components mixture of a Dirac mass, a Gaussian distribution and a Skew-Normal distribution.
According to the specified model, the age-at-death distribution is characterized via seven parameters corresponding to the main aspects of infant, adult and old-age mortality.
The proposed approach focuses on coherent modeling of multiple countries, and following a Bayesian approach to inference we allow to borrow information across populations and to shrink parameters towards a common mean level, implicitly penalizing diverging scenarios.
Dynamic modeling across years is induced through an hierarchical dynamic prior distribution that allows to characterize the temporal evolution of each mortality component and to forecast the age-at-death distribution.
Empirical results on multiple countries indicate that the proposed approach outperforms popular methods for forecasting mortality, providing interpretable insights on its evolution. 
\end{abstract}

{\small {\bf Keywords:}  Bayesian inference; Dynamic modeling; Mixture model; Skew-Normal distribution.}
\section{Introduction}
\label{sec:intro}
\begin{figure}[t]
	\includegraphics[width=\textwidth]{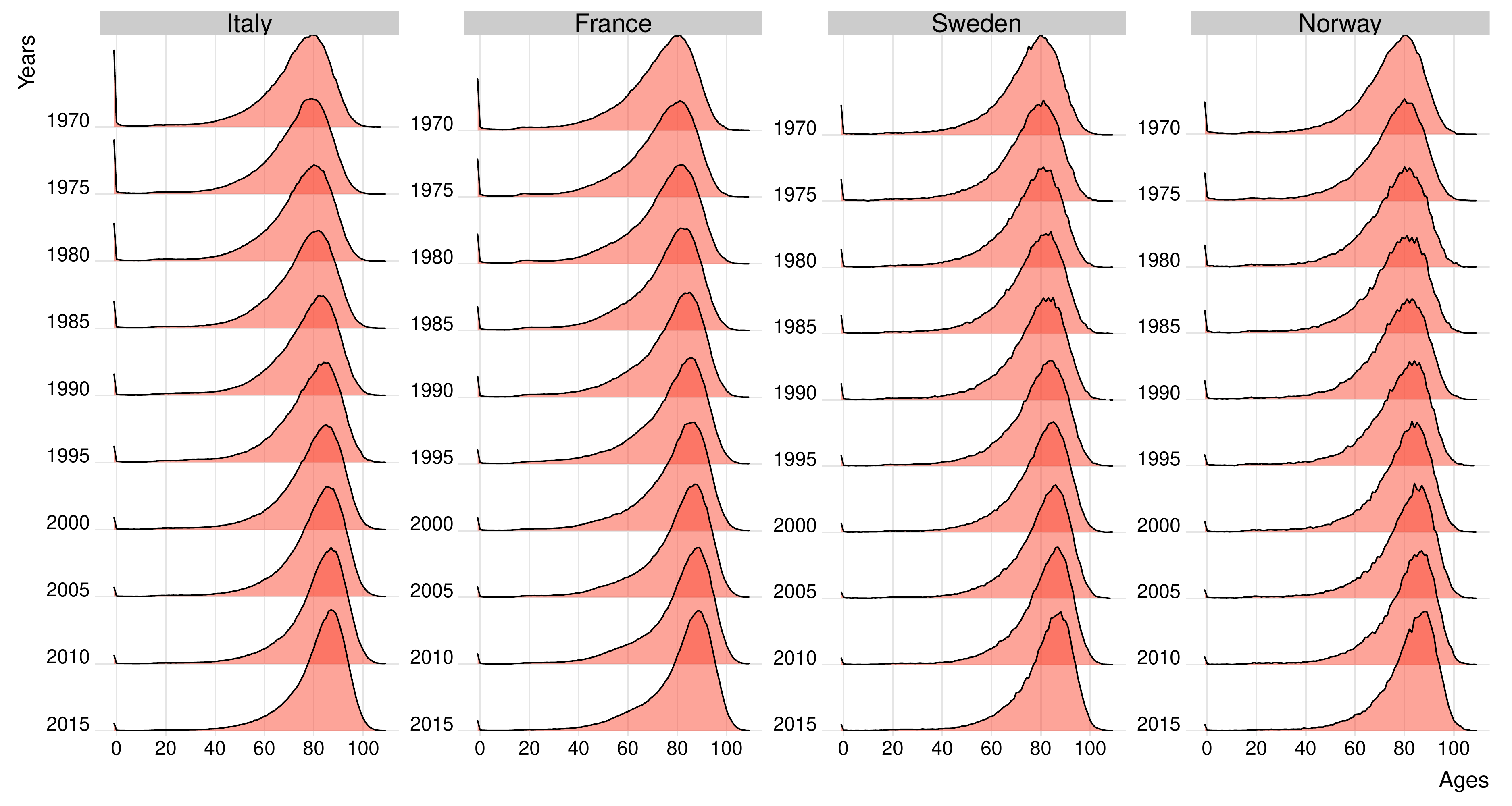}
	\caption{Age-at-death distribution across Italy, France, Sweden and Norway from $1970$ to $2015$. }
	\label{fig:data}
\end{figure}
In the last decades, changes in life expectancy and population growths have stimulated growing interest in sophisticated models for mortality \citep[e.g.,][]{Booth:2020,zanotto:2020}. 
These approaches characterize the observed age-specific regularities of the mortality function and their trends over time,  providing a parsimonious representation of the components of mortality and allowing to provide forecasts.
Estimates of past mortality trends provide insights on the global health of a population, describe relevant aspects of its demographic structure and facilitate comparisons across countries.
Similarly, probabilistic forecasts are routinely employed by researchers, practitioners and organizations to aid and support welfare planning strategies such as pension funds, insurances and health systems, among many others \citep[e.g.,][]{lutz:2010}.

Human mortality has been generally characterized by several quantities, with the hazard function, the survival function and the probability density function being the most popular ones \citep[e.g.,][]{klein:2006}.
Although each representation can be directly obtained from the others, their interpretations are complementary and highlight more clearly different aspects of mortality \citep[e.g.,][]{basellini:2019}.
In this article we will focus on modeling the distribution of the ages at death, generally referred to as ``age-at-death distribution'' and corresponding to the density function that characterizes mortality  \citep{keyfitz:2005}.
In the literature on modeling and forecasting mortality, this quantity has received significantly less attention than other quantities such as mortality rates \citep[e.g.,][]{lee:1992,li:2005}.
However, as outlined in \citet{basellini:2020} and \citet{pascariu:2019}, analysis of the age-at-death distribution can be incredibly useful, since it provides a direct evaluation of longevity and lifespan variability that cannot be directly inferred from other quantities \citep[see also][]{cheung:2005,canudas:2010}.

To further see that, \cref{fig:data} illustrates the age-specific distribution of deaths from four illustrative countries, over a time period that ranges from $1970$ to $2015$. 
These data highlight  crucial aspects of the temporal evolution of mortality, such as the global shift of the curve to older ages---due to the increase in life expectancy---and the drop of perinatal and infant mortality.
\cref{fig:data} also suggests that such an evolution has characterized countries differently, resulting in faster changes in some countries (e.g., Italy) and slower in others (e.g., Sweden).
Furthermore, the age-specific distribution of deaths illustrates that several mortality trends are converging,   resulting in similar patterns in terms of modal age at death, infant mortality and compression of old age mortality.

The first approaches for modeling the distribution of deaths date back to the 19-th century \citep[e.g.,][]{gompertz:1825,makeham:1860,pearson:1897}, and have been generalized in the following century \citep{siler:1983,heligman:1980}.
The main interest of such methods has been on characterizing mortality patterns across ages, focusing on a given country during a specific year. 
According to these approaches, the mortality curve can be characterized using different parametric specifications, accounting for distinct aspects of the mortality curve; see also \citet{dellaportas:2001} and \citet{mazzuco:2018}.
One notable limitation of these approaches is the focus on single-country modeling; when interest is on forecasting mortality for multiple countries, independent country-specific analysis are required.
This strategy ignores the fact that in most developed nations similar trends are observed, and this abundance of information can be appropriately included in the model to improve the quality of the estimates.
Joint modeling of multiple nations, in fact, is expected to improve the estimation of country-specific parameters and temporal trends, allowing to share information across different countries.

More recently, researchers have focused on explicitly modeling the temporal evolution of the age-at-death distribution, forecasting the parameters using time-series specifications \citep{pascariu:2019,basellini:2019}. 
These methods borrow ideas from death-rates forecasting, such as the milestone approach introduced by \citet{lee:1992} and sequent developments \citep[e.g.,][]{hyndman:2007}. 
Potentially, these models can be both used to describe the evolution of mortality in the past years and to provide forecasts, although the focus of applications is more frequently on forecasting only.
However, separate forecasts often lead to unreasonable results, with country-specific or sex-specific predictions that diverge substantially across groups.
Such divergent forecasts across sub-populations are implausible, since industrialized countries have been showing global convergence trends in terms of mortality levels,  and this tendency is expected to endure \citep{wilson:2001,bergeron:2017,oeppen:2008}.

Following the breakthrough approach of \citet{li:2005}, different authors have focused on ``coherent'' mortality forecasts, defined as  predictions that do not diverge across groups.
Such a strategy has constantly received attention in the past years, being successfully implemented in several models for death rates; for example, generalizing compositional-data techniques \citep{bergeron:2017} and functional-data models \citep{hyndman:2013} for mortality.
However, a critical issue with these methods concerns with the choice of a population standard to be used as a reference.
This choice is substantially arbitrary, and a common consensus on which sub-population should be used is still lacking, since results are often sensitive to different references \citep{Booth:2020,kjaergaard:2016}. 

Motivated by the above discussion, in this article we propose a novel approach for modeling and forecasting the age-at-death distributions of multiple countries.
We will focus on the age-distribution of deaths, combining the life tables quantities with the total number of deaths. 
This procedure allows to naturally take into account the population size of each country and its age-structure, which are important determinants of the age-at-death distribution.
The proposed methods can also be applied, without modifications, directly to the life table quantities, where we can assume that data are provided on artificial populations with fixed population size.
However, this would discard an important source of uncertainty, which corresponds to the population size of a specific nation.
Indeed, imposing a fixed same sample size weights each nation equally, implying that a relatively small country (e.g. Sweden) would have the same influence as a much bigger country (e.g., Germany or United Kingdom).
Therefore, we prefer to use the actual sample size of each county to have a proper quantification of uncertainty and genuinely account for different population structures.

We propose a method based on a mixture distribution function of a Dirac mass, a Gaussian distribution and a Skew-Normal distribution, in order to characterize infant, adult and old-age mortality with an interpretable set of parameters.
Following a Bayesian approach to inference, we specify hierarchical priors for the dynamic parameters and the country-specific coefficients, relying on informative common prior specifications to improve the borrowing of information across nations.
Compared to the available techniques, our method allows to model directly the temporal evolution and the country-specific variations across multiple sub-populations, avoiding sequential procedures and providing estimates shrunk towards common levels.
Under the proposed methods, coherent predictions are naturally obtained without selecting a reference population or an external standard, but instead estimating these quantities from the available data.
We will refer to the proposed dynamic model based on a skewed distribution function as \name{} in the sequel.

\section{Methods}
\subsection{Model specification}

\begin{figure}[t]
	\centering
	\includegraphics[width=.8\textwidth]{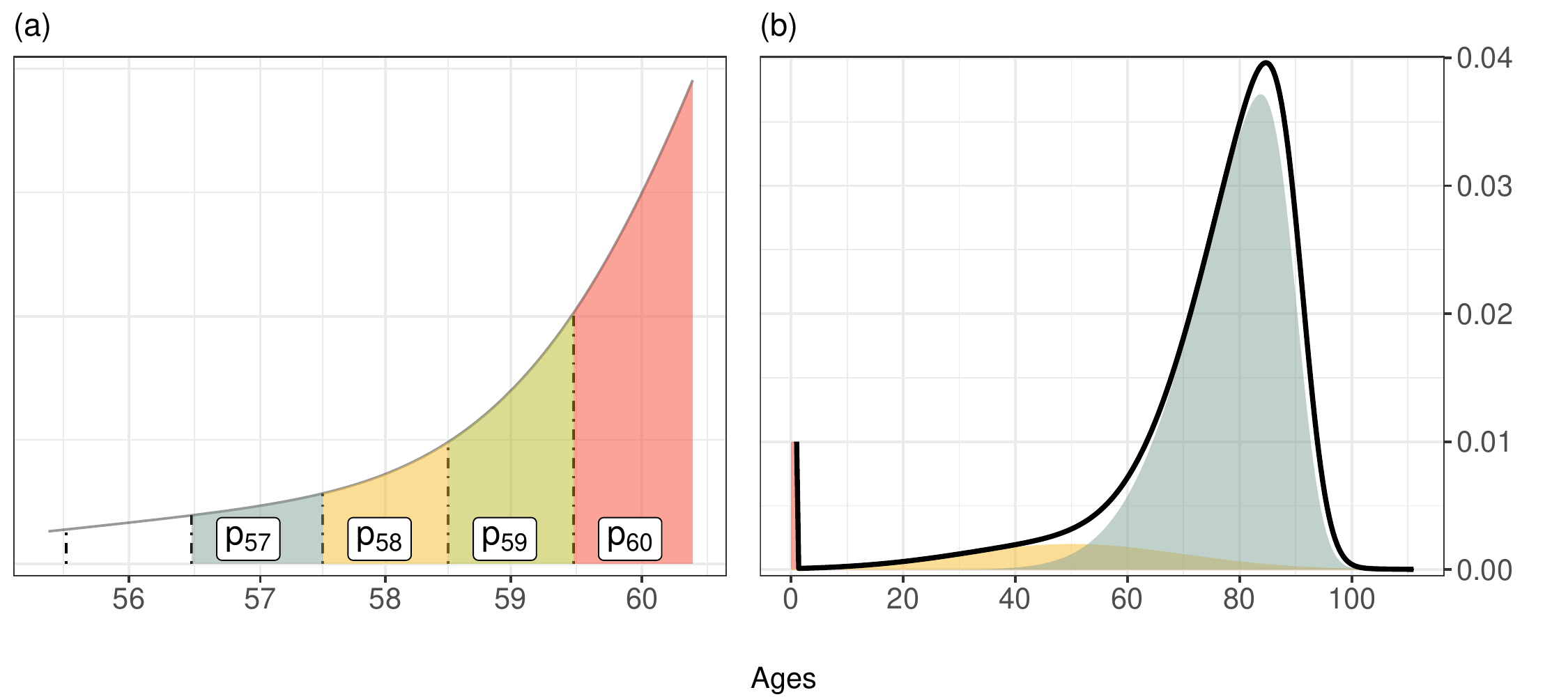}
	\caption{Proposed mixture density function.
Panel (a) provides a sketch of the discretization of the \textsc{pdf} into a set of probabilities $p_j$, with  $j = 57, \dots, 60$.
Panel (b) highlights the three components of the mixture density function.}
	\label{fig:mix}
\end{figure}

Our interest is on providing a flexible representation of the age-at-death distribution in multiple countries, dynamically across different years or time periods.
We will characterize this quantity through a continuous probability density function (\textsc{pdf}) $f(x; \boldsymbol{\vartheta}_{jt})$ which parametrizes the probability of dying at age $x$ -- based on the individuals current living at age $x$ -- in country $j$ during year $t$
via a set of country- and time-specific parameters $\boldsymbol{\vartheta}_{jt}$, with $x=0,\dots,110$, $j=1,\dots,p$ countries and $t=1,\dots,T$ years.
Although $f(x; \cdot)$ is often specified as a continuous function, in practice we focus on characterizing the age-at-death distribution on a discrete grid, corresponding to the probability $p_{x\,jt}$ of dying at a specific age $x=0,\dots,110$.
Such an aim can be addressed discretizing $f(x; \boldsymbol{\vartheta}_{jt})$ over a set of disjoint intervals $(x - 1/2, x+1/2]$ as
\begin{equation}
	\label{eq:int}
	p_{x\,jt} = \int_{x-1/2}^{x+1/2} f(z; \boldsymbol{\vartheta}_{jt}) \,\dif z = F(x+1/2; \boldsymbol{\vartheta}_{jt}) -  F(x-1/2; \boldsymbol{\vartheta}_{jt}), \quad x=0,\dots,110,
\end{equation}
where $F$ denotes the cumulative distribution function (\textsc{cdf}) associated with $f$.
This characterization of probabilities via a discretization of an underlying \textsc{pdf} automatically
guarantees that the vector $\mathbf{p}_{jt} = \mathbf{p}_{jt}(\boldsymbol{\vartheta}_{jt}) = [p_{0\,jt}, \dots, p_{110\,jt}]$ with elements as in Equation~\ref{eq:int} is a proper probability vector with ${p_{x\,jt} \in [0,1]}$ and ${\sum_x p_{x\,jt} = 1}$, therefore avoiding further restrictions \citep[e.g.,][]{oeppen:2008}; see the left panel of \cref{fig:mix} for a graphical representation of the discretization process.
We denote as ${\mathbf{D}_{jt} = [D_{0\,jt}, \dots, D_{110\,jt}]}$ the number of deaths and as $\mathbf{p}_{jt} = [p_{0\,jt}, \dots, p_{110\,jt}]$ the death probabilities, where elements $D_{x\,jt}$ and $p_{x\,jt}$ denote, respectively, the number of deaths and the probability of dying at age $x$ -- based on the current individuals living at age $x$ -- in country $j$ during year $t$.
Given the total number of deaths ${n_{jt} = \sum_x D_{x\,jt}}$ and the probabilities outlined in \cref{eq:int}, we model the vector $\mathbf{D}_{jt}$ as a multinomial distribution with 
\begin{equation}
	(\mathbf{D}_{jt} \mid n_{jt}, \mathbf{p}_{jt}) \sim \mbox{\textsc{multinom}}(n_{jt}, \mathbf{p}_{jt}), \quad j = 1, \dots, p, \quad t = 1, \dots, T.
	\label{eq:multi}
\end{equation}

\noindent
A crucial aspect of this approach involves the specification $f(x; \cdot)$, which induces a model on the vector of probabilities $\mathbf{p}_{jt}$.
In practice, different parametric models for $f(x; \cdot)$ are available in the literature, with the \citet{gompertz:1825} and \citet{siler:1983} models being popular options. 
The age-at-death distribution $f(x; \cdot)$ could be also modeled semi-parametrically, using for example splines or other basis functions \citep{hyndman:2013,basellini:2020}. 
However, when interest is explicitly on characterizing relevant aspects of mortality -- such as the impact of infant mortality or the shape of old-age mortality -- parametric models provide a much clearer interpretation and deeper insights on the structure of the populations under investigation.
In addition, mortality can be naturally divided into different components, accounting for deaths at different phases of life \citep[e.g.,][]{lexis:1879,pearson:1897}.
Therefore, it is desirable to rely on a specification which preserves such a decomposition, for example using mixtures of distributions \citep{carriere:1992,mazzuco:2018}.

In this work, we draw inspiration from \citet{mazzuco:2018} and \citet{zanotto:2020} and rely on a mixture of three components to characterize mortality; see also \citet{carriere:1992} for related arguments.
Specifically, in order to facilitate model's interpretation, we decompose mortality as the contribution of infant mortality, adult mortality and old-age mortality. 
Infant mortality is accounted through a single Dirac mass at age $0$, while adult mortality is modeled via a Gaussian distribution with mean $\mu_{jt} \in \mathbb{R}$ and standard deviation $\sigma_{jt} \in \mathbb{R}^{+}$.
Old-age mortality, instead, is characterized using a Skew-Normal distribution \citep[e.g.,][]{azza:2013} with location parameter $\xi_{jt} \in \mathbb{R}$, scale parameter $\omega_{jt} \in \mathbb{R}^{+}$ and shape parameter $\alpha_{jt} \in \mathbb{R}$. 
The impact of each component to the overall mortality function is accounted assigning mixture weights $\pi_{h\,jt} \in [0,1]$ such that ${\pi_{0\,jt} + \pi_{1\,jt} + \pi_{2\,jt} = 1}$ in order to guarantee that the resulting expression remains a proper \textsc{pdf}.
This choice leads to the following characterization of $f(x; \boldsymbol{\vartheta}_{jt})$
\begin{align}
	\label{eq:mixf}
	f(x; \boldsymbol{\vartheta}_{jt}) =\pi_{0\,jt} \mathds{1}[x=0] + \pi_{1\,jt}\frac{1}{\sigma_{jt}}\phi\left(\frac{x - \mu_{jt}}{\sigma_{jt}}\right)  +
									 \pi_{2\,2t} \frac{2}{\omega_{jt}}\phi \left(\frac{x - \xi_{jt}}{\omega_{jt}}\right) \Phi\left(\alpha_{jt}\frac{x - \xi_{jt}}{\omega_{jt}} \right),
\end{align}
\noindent
with $\boldsymbol{\vartheta}_{jt} = \left(\pi_{1\,jt}, \pi_{2\,jt}, \mu_{jt}, \sigma_{jt}, \xi_{jt}, \omega_{jt}, \alpha_{jt} \right)$, $\mathds{1}[\cdot]$ denoting the indicator function and $\phi(\cdot)$, $\Phi(\cdot)$ denoting the \textsc{pdf} and the \textsc{cdf} of a standard Gaussian, respectively; we omit the parameter $\pi_{0\,jt}$ from $\boldsymbol{\vartheta}_{jt}$ since $\pi_{0\,jt} = 1- \pi_{1\,jt}-\pi_{2\,jt}$ due to the constraint on the mixture weights; see the right panel of \cref{fig:mix} for a graphical representation of the proposed model for the age-at-death distribution.

\cref{eq:mixf} allows to characterize the age-at-death distribution with flexibility, while also retaining interpretability of the mortality structure.
The first component of \cref{eq:mixf} accounts for infant and perinatal mortality, characterizing deaths in the first year of life.
As outlined, for example, in \cref{fig:data}, perinatal mortality is notably decreased during recent years, mainly due to technological advances \citep[e.g.,][]{wilson:2001,zanotto:2020}; the mixture weight $\pi_{0\,jt}$ characterizes the impact of this component in the overall mortality.
The use of a Dirac mass is particularly adequate for developed countries, where infant mortality occurs during the first year of life and can be properly modeled with a single mass, avoiding more elaborate specifications and additional parameters.
However, \name{} can be naturally adapted to countries with high infant-mortality levels, replacing the Dirac mass with an half-normal distribution \citep{zanotto:2020} and leveraging on the same model architecture; in \cref{sec:addSpec}, we compare such a specification with the proposed model.
The Gaussian component, instead, is assigned to a relative weight $\pi_{1\,jt}$ and characterizes adult mortality via the location parameter $\mu_{jt}$ and the scale parameters $\sigma_{jt}$.
In addition, these quantities correspond also to the moments of the Gaussian component, and therefore they can be interpreted directly as the mean and standard deviation of the age at adult mortality.

The Skew-Normal component, lastly, accounts for the asymmetric shape of old-age mortality, which has been frequently observed in developed countries \citep[e.g.,][]{mazzuco:2018,pascariu:2019}.
Since most deaths occur at old ages, this component is expected to be associated to the largest mixture weight $\pi_{2\,jt}$.
Inference on the structure of old-age mortality rely on such mixture weight and on the Skew-Normal parameters $\xi_{jt}, \omega_{jt}$ and $\alpha_{jt}$, providing insights on the evolution of old-age mortality in terms of location, scale and shape of such a component.
In practice, it can be convenient to focus on the mean and standard deviation of old-age mortality, which are more interpretable measures of the shift and compression of the age-at-death distribution.
According to the proposed parametrization, these quantities can be conveniently expressed as 
\begin{equation}
	\tilde{\mu}^{\mbox{ \tiny SN }}_{jt} =  \xi_{jt} +\omega_{jt} \delta_{jt} \sqrt{2/\pi}, \quad
\tilde{\sigma}^{\mbox{ \tiny SN }}_{jt} =\omega_{jt}\sqrt{1-2\delta^{2}_{jt}/\pi} \quad 
\label{eq:sn_mom}
\end{equation}
with $\delta_{jt} = \alpha_{jt} / \sqrt{1 + \alpha_{jt}}$ \citep[e.g.,][sec. 2.1.4]{azza:2013}.
Alternatively, it is also possible to parameterize directly the Skew-Normal distribution in terms of its cumulants  \citep[e.g.,][sec. 3.1.4]{azza:2013}.  However, for simplicity in implementation and exposition, we prefer to focus on the outlined parameters, and eventually post-process the estimates to focus on the functionals of interest.

\subsection{Hierarchical dynamic model}

The Multinomial model outlined in Expression~\ref{eq:multi}, with probabilities given by \cref{eq:int,eq:mixf}, provides a flexible specification for the distribution of the number of deaths occurred in a single country $j$ during a specific year $t$.
In order to account for the temporal variation of the age-at-death distribution, we model explicitly the dynamic evolution of the parameters $\boldsymbol{\vartheta}_{jt}$ through an hierarchical dynamic model.
This strategy will allow us to highlight what aspects of mortality have changed in the past years, characterizing the rates of such changes and facilitating comparison across different countries.
In addition, the inclusion of a dynamic component facilitates the developments of mortality forecasts extrapolating parameters in time, obtaining predictions for the future age-at-death distributions.

We specify the hierarchical model in an unconstrained domain, transforming the parameters of \cref{eq:mixf}.
This choice facilitates computation and implementations, allowing to rely on Gaussian dynamic models in the transformed space and avoiding restrictions on the support of the parameters.
For this purpose, we introduce the reparametrized vector ${\boldsymbol{\tilde{\vartheta}}}_{jt} $, with elements  $(\tilde{\vartheta}_{jt \,1}, \dots,  \tilde{\vartheta}_{jt \,7})$ corresponding to
\begin{equation}
	\widetilde{\boldsymbol{\vartheta}}_{jt} = (\tilde{\vartheta}_{jt \,1}, \dots,  \tilde{\vartheta}_{jt \,7})\coloneqq \left(\log\left(\frac{\pi_{1\,jt}}{1-\pi_{1\,jt}}\right), \log\left(\frac{\pi_{2\,jt}}{1-\pi_{1\,jt}-\pi_{2\,jt}}\right), \mu_{jt}, \log(\sigma_{jt}), \xi_{jt}, \log(\omega_{jt}), \alpha_{jt} \right).
	\label{eq:repar}
\end{equation}
In \cref{eq:repar}, the mixture weights $(\pi_{1\,jt}, \pi_{2\,jt})$ are mapped into $(\tilde{\vartheta}_{jt\,1}, \tilde{\vartheta}_{jt\,2})$ leveraging a cumulative logit transformation, while $\sigma_{jt}$ and $\omega_{jt}$ -- corresponding to adult and old-age mortality scale parameters -- are transformed into $\tilde{\vartheta}_{jt\,4}$ and $\tilde{\vartheta}_{jt\,6}$, respectively, via a logarithmic transformations.
Note that each $\tilde{\vartheta}_{jt \, k} \in \mathbb{R}$ for $k=1,\dots,7$, and that the vector $\widetilde{\boldsymbol{\vartheta}}_{jt}$ effectively corresponds to a one-to-one reparametrization of ${\boldsymbol{\vartheta}}_{jt}$.

We model the dynamic evolution of the elements $\tilde{\vartheta}_{jt \, k}$ leveraging a random-walk model with drift. This choice leads to the following specification. 
\begin{equation}
\begin{split}
	\tilde{\vartheta}_{j0 \, k} &\sim \mathcal{N}(m_{k}, s_{k}^2) \\
	(\tilde{\vartheta}_{jt \, k} \mid \tilde{\vartheta}_{j\,t-1\,k}, \beta_{j\,k}, \eta_{j\,k}^2) &\sim \mathcal{N}(  \beta_{j\,k} +  \tilde{\vartheta}_{j\,t-1\,k}, \eta_{j\,k}^2), \quad t = 1, \dots, T,
\end{split}
\label{eq:ts_mod}
\end{equation}
In \cref{eq:ts_mod},  $\tilde{\vartheta}_{j0 \, k} \in \mathbb{R}$ denotes the initial condition of $\tilde{\vartheta}_{jt \, k}$, and it is assigned to a Gaussian distribution with mean  $m_{k}  \in \mathbb{R}$ and standard deviation $s_{k} \in \mathbb{R^+} $.
These values correspond to the initial information for the dynamic hierarchical model, and characterize a probabilistic representation of our beliefs before the data are observed \citep[e.g.,][]{west:DN}. 
The drift parameter $\beta_{j\,k} \in \mathbb{R}$ measures the expected difference across consecutive values of $\tilde{\vartheta}_{jt \, k}$ and accounts for possible increasing or decreasing linear trends in the evolution of the parameters.
The inclusion of this component is particularly relevant in our problem, where we expected to observe clear and important trends in the evolution of mortality, accounted by the parameters charactering the age-at-death distribution.
Conditionally on the value at time ${(t-1)}$, each $\tilde{\vartheta}_{jt \, k}$ is recursively distributed as a Gaussian distribution with mean $ \beta_{j\,k} + \tilde{\vartheta}_{j\,t-1\,k}$, and standard deviation $\eta_{j\,k} \in \mathbb{R^+}$.
Overall, we can interpret \name{} as a Multinomial state-space model, where \Cref{eq:ts_mod} controls the evolution of the latent states via random-walks with Gaussian innovations and, conditional on the states, the observed vectors of deaths are modeled as independent Multinomial variables with probabilities given by \cref{eq:mixf,eq:int}; see \cref{dagM} for a graphical representation of the proposed hierarchical dynamic model.

Although this dynamic model might seem oversimplistic, we found that such a specification leads to good results in a large variety of applications involving developed countries; see \cref{sec:app} for further details.
More elaborate extensions are possible, for example including a moving-average component, increasing the lag of the autoregressive terms or joint modeling multiple components of $\tilde{\vartheta}_{jt \, k}$ via multivariate random-walks.
However, a random walk with drift on the latent states often characterizes well many phenomena without over-parameterizing the dynamic component, and is also general enough to be directly used, without further modifications, in a large variety of settings; see \citet{durbin:2012} and \citet{west:DN} for related arguments and \cref{sec:addSpec} for a comparison with alternative specifications.
 
Additionally, we highlight that \cref{eq:ts_mod} is valid for any equally-spaced time grid, and therefore the proposed approach can be also applied to different time scales.
In fact, our description focuses on yearly data for simplicity in exposition, but this requirement is not necessary for developing the proposed model in practice.
Several important settings involve analysis of mortality across different time scales, such as monthly data or multi-year periods; for example, when interest is on modeling short-term variations due to extraordinary events (e.g., \textsc{covid-19} pandemic) or large-scale trends, respectively.
The general structure of \name{} facilitates its use in these different settings, without modifying the model specification.

\input{dag.tex}

\subsection{Bayesian inference}

As discussed in \cref{sec:intro}, a crucial aspect of mortality forecasting involves the penalization of diverging scenarios via coherent modeling.
This can be obtained explicitly including a common trend across countries \citep[e.g.,][]{li:2005,li:2013}, or choosing a reference population \citep[e.g.,][]{hyndman:2013,basellini:2019}; see also \citet{Booth:2020} for a recent discussion.
However, these choices are arbitrary and significantly affect the final results \citep[e.g.,][]{kjaergaard:2016}.
Therefore, we follow a different path and induce coherent predictions via a Bayesian approach to inference, regularizing estimates toward a common mean trend.
More specifically, we introduce a common prior specification for the country-specific dynamic hierarchical model defined in \cref{eq:ts_mod}, for each coefficient $k=1,\dots,7$.
Including the prior for $\tilde{\vartheta}_{j0 \, k}$, we let
\begin{equation}
	\begin{split}
		\tilde{\vartheta}_{j0 \, k} &\sim \mathcal{N}(m_{k}, s_{k}^2) \\
		\beta_{j\, k} \simI \mathcal{N} (m_{\beta_k}, s_{\beta_k}), &\quad \eta_{j\,k}^2 \simI \mbox{Inv-Gamma} (\mbox{a}_{k}, \mbox{b}_k) \quad \mbox{for} \,\, j = 1, \dots, p,
		\label{eq:prior}
	\end{split}
\end{equation}
where $m_{\beta_k}, s_{\beta_k},  \mbox{a}_k$ and  $\mbox{b}_k$ denote fixed hyper-parameters;  see \cref{dagM2}  for an illustration of the hierarchical dependence structure induced by the common prior specification.

The specification of a common prior distribution for the country-specific parameters allows to share information across different nations, improving estimates for the age-at-death distributions.
In addition, the introduction of a common prior shrinks estimates toward a common mean, providing an implicit regularization on the parameters.
Such a shrinkage is what makes \name{} a ``coherent'' model, using a data-driven reference mortality level without the need of a user-defined benchmark.

\input{dag2.tex}

\subsection{Prior elicitation and posterior computation}

The large abundance of research on the evolution of mortality, and the clear interpretation of the parameters in \cref{eq:mixf}, facilitate informative elicitation for the hyper-parameters of \cref{eq:prior}.
For example, European adult mortality is expected to lie between $30$ and $70$ years with high probability \citep[e.g.,][]{lewer:2020}; therefore, we can assume $\tilde{\vartheta}_{j0\,3} \sim \mathcal{N}(50, 10)$ for the initial condition of $\tilde{\vartheta}_{jt \, 3}$, independently for $j=1,\dots,p$.
Similarly, informative specifications for the prior distribution on the drift parameter $\beta_{j\,k}$ allow to include information on the direction and intensity of the temporal trends.
For instance, we could center the trend $\beta_{j\,3}$ around $m_{\beta_3} = 2$, resulting in an expected increasing trend for the age at adult mortality of $2$ years between consecutive time points.
In \cref{sec:addSpec}, we compare such informative prior specification with a non-informative elicitation, confirming the importance of the proposed approach to induce shrinkage across countries and obtain coherent predictions.

There is a large literature on simulation methods for non-Gaussian state-space models; see, for example, \citet[Chapter 13.3]{durbin:2012},  \citet[Chapter 6]{prado:2010} or \citet{chopin:2020} for recent developments.
A major complexity in our methods derive from the computation of the age-at-death distribution, which links ${\boldsymbol{\vartheta}}_{jt}$ to $\mathbf{p}_{jt}$ via \cref{eq:int}.
Therefore, we conduct posterior inference relying on the \textsc{r} package \texttt{nimble} \citep{nimblePack}, that allows to specify the main structure of \name{}, compile it in \textsc{c++}, and implement an adaptive Metropolis-within-Gibbs \textsc{mcmc} algorithm to simulate from the posterior distribution of the model's parameters.
Specifically, the parameters of \cref{eq:prior} are updated using Gibbs steps, while the remaining are updated with multivariate Metropolis steps, due to the lack of conjugacy induced by \cref{eq:int}.
Pseudo-code illustrating the main steps to conduct posterior inference is reported in the Supplementary materials.
Forecasts for future time points $t > T$ can be simulated recursively from the posterior predictive distribution, continuing the conditional representation of \cref{eq:ts_mod} for each draw of the \textsc{mcmc} output. 

In order to improve the computational performance, we included further modifications to efficiently compute the \textsc{cdf} of the mixture density function in Expression~(\ref{eq:mixf}) using \textsc{c++}.
This operation is non trivial since it requires to compute the \textsc{cdf} of the Skew-Normal component, which relies on the \citet{owen:1956} T-function; see also \citet[Appendix B]{azza:2013} for the details and \citet{patefield:2000} for practical considerations.
Overall, posterior computations in a setting with $p= 12$ countries and $T=20$ time points requires roughly $2$ minutes per $1000$ iterations on an $8$-core \textsc{intel} {\small I{$7$-$7700$}HQ}, $2.8$ \textsc{ghz} processor running Linux, which is satisfactory for our purposes.

\section{European countries mortality}
\label{sec:app}
\subsection{Data description}
The focus of this Section is to evaluate the ability of \name{} to characterize the age-at-death distributions of different European countries.
In \cref{sec:appwin}, we are interested in comparing its performance with some state-of-the-art methods for mortality forecast, focusing on rolling windows scenarios.
In \cref{sec:appeu} we use the proposed method to estimate mortality and to forecasts for future years, 
illustrating in details how \name{} can be used to draw thoughtful inferential conclusions by making inference on different quantities.
Analysis will focus on the time range $1960-2016$, and on joint modeling the following $p=12$ countries: Austria (\textsc{aut}),    Switzerland (\textsc{che}),    England \& Wales (\textsc{gbr}), Germany (\textsc{deu}),  Denmark (\textsc{dnk}),    Finland (\textsc{fin}),    France (\textsc{fra}),    Italy (\textsc{ita}),    Netherlands (\textsc{nld}),    Norway (\textsc{nor}),    Spain (\textsc{esp}) and    Sweden (\textsc{swe}), separately for the male and female populations.
These countries were selected according to the European countries analyzed in other approaches, such as \citet{li:2005}.
Mortality data are retrieved from the \citet{hmd}, and the year- and country-specific number of deaths are obtained with the utilities of the \textsc{r} package \texttt{demograpy} \citep{demo:R}.

\subsection{Performance evaluation via rolling window}
\label{sec:appwin}

In order to quantitatively evaluate the performance of \name{} in estimating and forecasting mortality, we rely on rolling window scenarios.
Specifically, analysis focus on sub-periods of $20$ years of data to fit the models, and subsequent periods of $10$ years of data to provide forecasts. 
Performance is evaluated both in terms of in-sample accuracy---comparing the model fit with the $20$ observed years---and in terms of forecasts ability---comparing predictions with the out-of-sample block for the subsequent $10$ years.
Multiple scenarios are obtained rolling the window forward of steps of $1$ years, thereby dividing the range $1960-2016$ in $30$ consecutive scenarios.

We compare \name{} with different popular approaches for mortality forecasting: the Lee-Carter model \citep{lee:1992}, the coherent Li-Lee model \citep{li:2005}, the functional mortality model of Hyndman–Ullah \citep{hyndman:2007}, the compositional Oeppen model \citep{oeppen:2008}, and the $\textsc{mem5}$ model \citep{pascariu:2019}.
All methods are conveniently available in a common interface within the \textsc{r} package \texttt{MortalityForecast} \citep{pascariu:R}. 
Since these methods focus on modeling single population data, we perform the analysis on each country separately. 
In order to compare the approaches on the same basis, results are evaluated in terms of the quality of the fitted and predicted age-at-death distributions and death rates.
Under \name{}, such estimates are obtained post-processing the \textsc{mcmc} output, computing the posterior distribution of \cref{eq:int} via Monte Carlo integration.
For the remaining approaches, we follow standard practices and transform predictions into death rates and age-at-death distributions \citep[e.g.,][]{pascariu:R}.

Posterior inference proceeds separately for the female and male populations, relying on informative prior distributions.
We elicited the initial conditions $\tilde{\vartheta}_{j0\,k}$ for the hierarchical dynamic component considering the expected level of mortality in the temporal window under investigation, centering old-age mortality around $70$ years ($m_5 = 70$), adult mortality around $50$ years $(m_3 = 50)$ and setting the remaining values of $m_k$ to $0$. 
We also let ${s_k} = 10$ for all $k$ to induce sufficient variability in such initial prior guesses and cover the temporal window under investigation. 
Lastly, we specified a shared standard Gaussian for the drift parameters $\beta_{j\,k}$ (setting $m_{\beta_k} = 0$ and $s_{\beta_k} = 1$) and a weakly-informative Inverse-Gamma on the variances of the dynamic innovations $\eta_{j\,k}^2$, setting $\mbox{a}_k = 0.01, \mbox{b}_k = 0.01$ for all $k$.
In each scenario, posterior inference for \name{} relies on $100000$ iterations collected after a burn-in of $5000$, thinning one iteration every $5$. Convergence was assessed in terms of mixing, autocorrelation of the chains and Geweke diagnostics, which resulted satisfactory in all the windows considered, with an Effective Sample Size larger than $18000$ out of $20000$ stored iterations for all the parameters.

Performance is evaluated in terms of Mean-Squared-Error (\textsc{mse}) and Mean-Absolute-Error (\textsc{mae}) between the fitted and predicted age-at-death distributions or death rates and the ground truth.
Since the elements of the age-at-death distribution are generally very small numbers, we report results in relative terms, dividing the performance of each approach by the performance obtained by \name{} in each scenario and country.
Recalling that both \textsc{mse} and \textsc{mae} are measure of discrepancy, results larger than $1$ indicate that predictions under \name{} achieve an error which is lower than the competitor, and therefore we can conclude that \name{} performs better; the opposite holds for values lower than $1$.

\begin{table}[tb]
\centering
\begin{tabular}{lcccc}
&	\multicolumn{2}{c}{\textsc{female}} &	\multicolumn{2}{c}{\textsc{male}}\\
\cmidrule{2-3}
\cmidrule{4-5}
	{\em Age-at-death} & \textsc{mae} & \textsc{mse} & \textsc{mae} & \textsc{mse} \\
 \toprule
	\name{} & 1.00                & 1.00                & 1.00                & 1.00 \\
Hyndman–Ullah   & 1.41 $[1.17, 1.64]$ & 2.73 $[1.91, 3.80]$ & 1.92 $[1.39, 2.40]$ & 5.37 $[2.76, 7.79]$ \\
Lee-Carter      & 1.40 $[1.17, 1.62]$ & 2.74 $[1.90, 3.79]$ & 1.93 $[1.39, 2.36]$ & 5.34 $[2.77, 7.77]$ \\
Li-Lee          & 1.23 $[0.94, 1.45]$ & 2.19 $[1.22, 2.97]$ & 1.75 $[1.30, 2.28]$ & 4.71 $[2.53, 6.99]$ \\
\textsc{mem5}   & 1.22 $[1.07, 1.31]$ & 2.01 $[1.55, 2.55]$ & 1.86 $[1.56, 2.19]$ & 4.20 $[3.01, 6.61]$ \\
 Oeppen         & 1.13 $[0.92, 1.36]$ & 1.88 $[1.31, 2.66]$ & 1.61 $[1.24, 2.14]$ & 3.79 $[2.24, 6.37]$ \\
 \bottomrule
\end{tabular}
\begin{tabular}{lcccc}
	 &&&&\\
	{\em Death rates} &&&&\\
 \bottomrule
	\name{} & 1.00                & 1.00                & 1.00                & 1.00 \\
Hyndman–Ullah   & 0.86 $[0.76, 0.95]$ & 1.02 $[0.86, 1.11]$   & 0.84 $[0.79, 0.92]$ & 1.02 $[0.85, 1.19]$ \\
Lee-Carter      & 0.91 $[0.79, 1.06]$ & 1.08 $[0.88, 1.86]$   & 0.87 $[0.80, 1.00]$ & 1.07 $[0.87, 1.58]$ \\
Li-Lee          & 1.02 $[0.71, 1.26]$ & 1.81 $[0.71, 5.42]$   & 0.99 $[0.77, 1.17]$ & 1.90 $[0.95, 3.89]$ \\
\textsc{mem5}   & 1.06 $[0.97, 1.14]$ & 0.99 $[0.90, 1.09]$   & 1.01 $[0.96, 1.05]$ & 1.02 $[0.95, 1.08]$ \\
 Oeppen         & 1.27 $[1.03, 4.71]$ & 3.87 $[1.10, 6.51]$ & 1.09 $[0.89, 2.98]$ & 1.32 $[0.81, 7.01]$ \\
 \bottomrule
\end{tabular}
\caption{In-sample relative performance in estimating the age-at-death distribution (top) and death rates (bottom), comparing predictions of \name{} against the competitors. Median across rolling windows and countries. First and third quartiles are reported in squared brackets.}
\label{app:iis} \end{table}

\begin{table}[tb]
\centering
\begin{tabular}{lcccc}
&	\multicolumn{2}{c}{\textsc{females}} &	\multicolumn{2}{c}{\textsc{males}}\\
\cmidrule{2-3}
\cmidrule{4-5}
	{\em Age-at-death} & \textsc{mae} & \textsc{mse} & \textsc{mae} & \textsc{mse} \\
 \toprule
	\name{} & 1.00                & 1.00                & 1.00                & 1.00 \\
Hyndman–Ullah   & 1.38 $[1.14, 1.69]$ & 1.99 $[1.39, 2.88]$  & 1.91 $[1.12, 2.48]$ & 3.64 $[1.42, 6.36]$ \\
Lee-Carter      & 1.30 $[1.03, 1.62]$ & 1.75 $[1.12, 2.65]$  & 1.73 $[1.06, 2.37]$ & 3.23 $[1.25, 5.54]$ \\
Li-Lee          & 1.28 $[0.91, 1.56]$ & 1.66 $[0.93, 2.54]$  & 1.73 $[1.06, 2.36]$ & 3.23 $[1.25, 5.53]$ \\
\textsc{mem5}   & 2.44 $[2.02, 3.03]$ & 5.67 $[3.98, 8.88]$  & 2.84 $[1.70, 3.88]$ & 7.92 $[3.23, 9.96]$ \\
 Oeppen         & 1.29 $[0.94, 1.82]$ & 2.30 $[1.29, 9.34]$ & 1.66 $[1.06, 2.36]$ & 3.86 $[1.45, 8.11]$ \\
 \bottomrule
\end{tabular}
\begin{tabular}{lcccc}
	 &&&&\\
	{\em Death rates} &&&&\\
 \toprule
	\name{} & 1.00                & 1.00                & 1.00                & 1.00 \\
Hyndman–Ullah   & 1.02 $[0.88, 1.10]$ & 1.35 $[1.18, 1.70]$   & 0.96 $[0.89, 1.05]$ & 1.24 $[1.05, 1.47]$ \\
Lee-Carter      & 1.12 $[0.95, 2.66]$ & 1.73 $[1.22, 8.96]$  & 1.06 $[0.94, 2.08]$ & 1.51 $[1.14, 9.24]$ \\
Li-Lee          & 1.12 $[0.95, 2.66]$ & 1.73 $[1.22, 7.95]$  & 1.06 $[0.94, 2.08]$ & 1.51 $[1.14, 9.24]$ \\
\textsc{mem5}   & 1.06 $[0.89, 1.19]$ & 1.03 $[0.76, 1.48]$   & 1.32 $[1.03, 1.58]$ & 1.05 $[0.83, 1.41]$ \\
 Oeppen         & 1.67 $[1.24, 4.07]$ & 3.99 $[1.53, 9.44]$ & 1.29 $[1.10, 1.69]$ & 1.97 $[1.36, 6.35]$ \\
 \bottomrule
\end{tabular}

\caption{Out-of-sample relative performance in estimating the age-at-death distribution (top) and death rates (bottom), comparing predictions of \name{} against the competitors. Median across rolling windows and countries. First and third quartiles are reported in squared brackets.}
\label{app:oos}
\end{table}

Results referring to in-sample evaluation are reported in \cref{app:iis}, computing the median, first and third quartiles of the relative performances across countries and rolling windows scenarios.
Current results suggest that \name{} outperforms the competitors both for the male and female populations, with a performance that is better than the other approaches in all settings.
These results confirm the ability of \name{} to characterize with flexibility age-at-death distributions across multiple countries and time periods, providing estimates that are consistent with the observed data and more accurate than state-of-the art methods. 
When \name{} is used for fitting death rates, a relatively worse performance is observed, particularly in terms of \textsc{mae}.
This result can be explained considering that old-ages $(80-100)$ are assigned more weight in the computation of death rates, compared to other quantities that characterize mortality \citep[e.g.,][]{keyfitz:2005}.
Therefore, \name{} performs particularly well in characterizing premature and adult mortality, and this leads to large improvements for modeling the age-at-death distribution, in particular for male populations where the share of adult mortality is relatively large.

Out-of sample performances are instead reported in \cref{app:oos}.
Results provide further evidence in favor of the ability of \name{}, which obtains out-of-sample forecasts more accurate than the competitors.
This important results suggest that, overall, \name{} is the most accurate methods for modeling and forecasting the age-at-death distribution in the $12$ European countries under investigation.
Performances stratified by countries are reported in the Supplementary materials.

\subsection{Results for European mortality}
\label{sec:appeu}

In \cref{sec:appwin}, we have demonstrated that \name{} outperforms the competitors in terms of the quality of the fitted and predicted age-at-death distributions.
In this Section we argue that the proposed method has concrete practical benefits in terms of mortality modeling also from an interpretative point of view.
Indeed, posterior inference on the parameters characterizing \name{} and their functionals provides insights on the structure of mortality and its future evolution, highlighting aspects that have larger impact in such changes and explaining differences across countries.

\begin{figure}[tb]
	\centering
	\includegraphics[width=\textwidth]{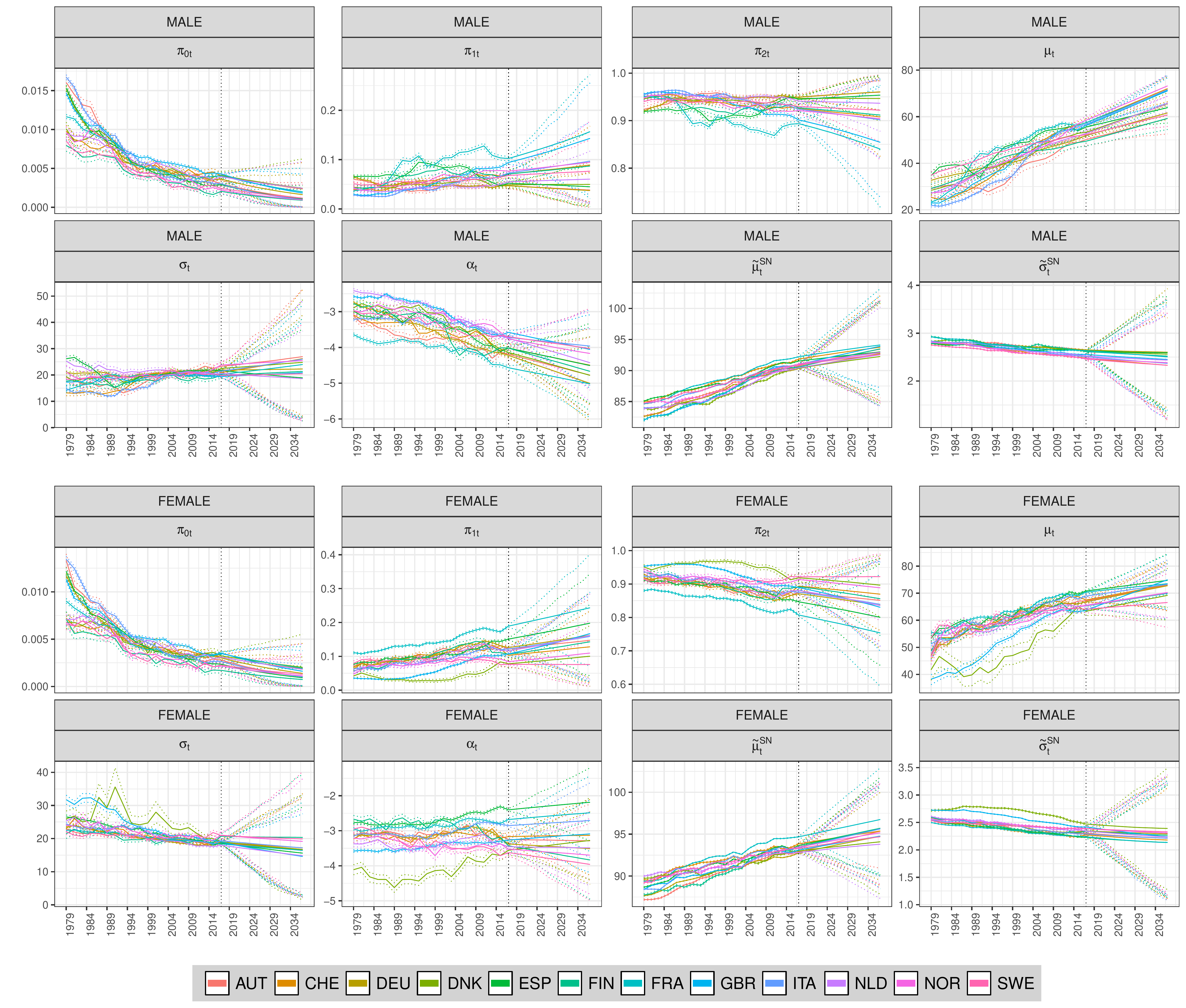}
	\caption{Posterior medians (solid lines) and $90\%$ credible intervals (dotted lines) for the parameters characterizing \name{} and functionals of interest. Black vertical dotted lines indicate the last observed period (2017).}
	\label{fig:parsM}
\end{figure}

To illustrate this, we  focus on the same European countries as in the previous section, modeling the more recent period $1980-2017$ and providing forecasts until $2037$.
Posterior simulation for this application relies on the same settings as in \cref{sec:appwin}.
Results in terms of mixing, autocorrelations, Geweke diagnostic and effective sample size were satisfactory for all the parameters considered, with an effective sample size larger than $19000$ out of $20000$ iterations for all the parameter.
We conduct inference on the set of parameters $(\pi_{0\,jt}, \pi_{1\,jt}, \pi_{2\,jt}, \mu_{jt}, \sigma_{jt}, \alpha_{jt}, \tilde{\mu}^{\mbox{ \tiny SN }}_{jt}, \tilde{\sigma}^{\mbox{ \tiny SN }}_{jt})$ to further facilitate interpretation.  Results are reported in \cref{fig:parsM}, illustrating the posterior medians and $90\%$ credible intervals for the estimated parameters for male and female populations separately.

Current empirical findings are consistent with known aspects of the temporal evolution of mortality of the past decades.
For example, the impact of infant mortality---measured via the parameters $\pi_{0j\,t}$---is notably decreased in all countries in the past years, dropping from values around $0.01$ to $0.005$.
Although both male and female populations share such decreasing trends, we also observe that infant mortality is larger for males than females \citep[e.g.,][]{drevenstedt:2008}.
Another clear aspect involves the evolution of the mean age of old-age mortality, captured via the Skew-Normal mean $\tilde{\mu}^{\mbox{ \tiny SN }}_{jt}$, which has constantly increased in all countries. 
The overall trend is similar between males and females, with woman reporting older ages on average.
Similarly, the decreasing trend of $\tilde{\sigma}^{\mbox{ \tiny SN }}_{jt}$ is consistent with the empirical evidence on compression of old-age mortality.

\begin{figure}[bt]
	\includegraphics[width=\textwidth]{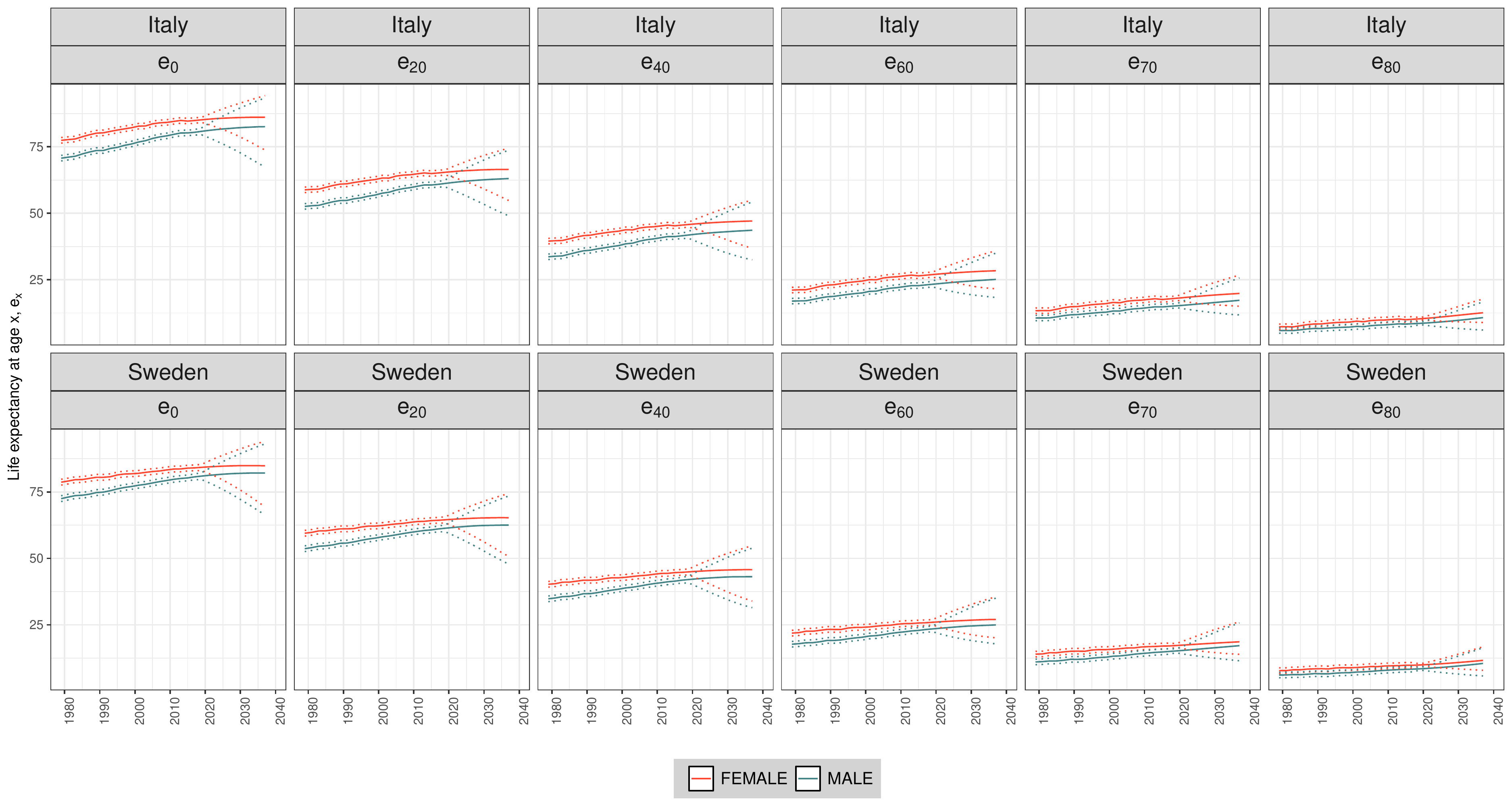}
	\caption{Life expectancy at ages $0$ (at birth), $20,40,60,70$ and $80$. Italy and Sweden, female and male population. Estimates are obtaining post-processing the \textsc{mcmc} for the age-at-death distribution. Solid and dashed lines denote posterior medians and $90\%$ credible intervals, respectively.}
	\label{fig:dataf}
\end{figure}

\begin{figure}[bt]
	\includegraphics[width=\textwidth]{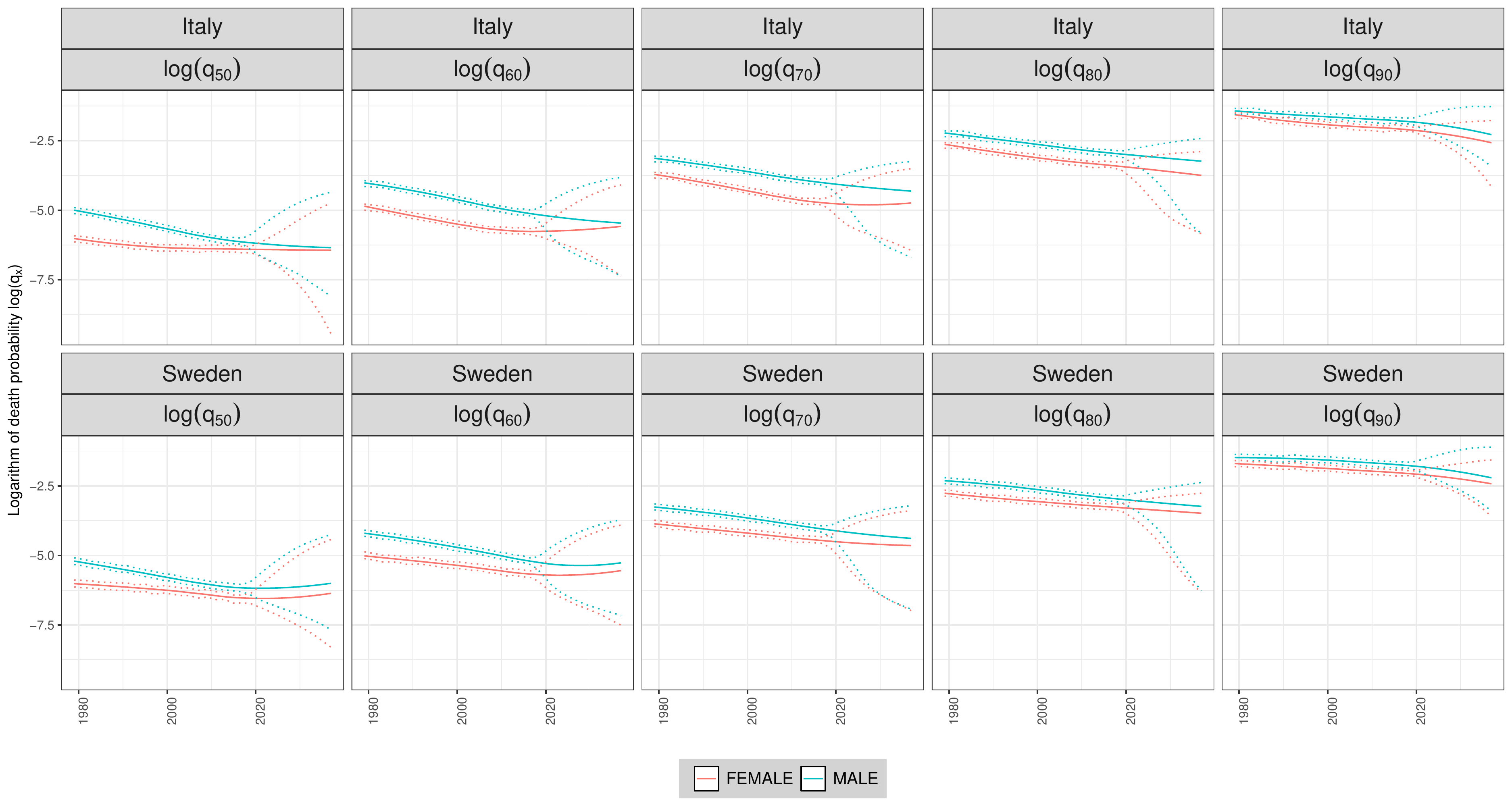}
	\caption{Logarithm of death probability at ages $50,60,70,80$ and $90$. Italy and Sweden, female and male population. Estimates are obtaining post-processing the \textsc{mcmc} for the age-at-death distribution. Solid and dashed lines denote posterior medians and $90\%$ credible intervals, respectively.}
	\label{fig:dataqx}
\end{figure}

Under \name{}, estimates and credible intervals for any complex functional of the model's parameters are easily obtained. 
More importantly, prediction intervals are determined taking into account the uncertainty of the entire inferential process, which is not the case for other coherent models, where all the uncertainty related to choice of reference is generally not accounted for.
Some interesting functionals include death rates and life expectancies, which can be obtained from the age-at-death distribution via deterministic transformations;
\name{} allows to conduct coherent inference also on this quantities, rigorously characterizing uncertainty post-processing the \textsc{mcmc} output. 
As an illustrative example, \cref{fig:dataf} and \cref{fig:dataqx} depict the life-expectancy and the death-probability, respectively, computed at different ages and considering male and female populations in Italy and Sweden.
Results are consistent with the conclusions outlines above, confirming regular patters of mortality such as non-overlapping trends for males and females and an overall improvement at all ages.

\subsection{Alternative parametric specifications}
\label{sec:addSpec}
In this section we evaluate the robustness of the parametric assumptions underlying \name{}, focusing on the application outlined in \cref{sec:appeu}.
We investigate different specifications for the mixture components characterizing \name{} and alternative prior specifications.
Specifically, we modify \name{} focusing on 
 \begin{itemize}[]
 	\item modeling old-age mortality with a rescaled Beta distribution, supported in the interval $[75-110]$, instead of a Skew-Normal distribution;
 \item modeling infant mortality with an Half-Normal distribution, as in \citet{zanotto:2020}, instead of a single Dirac mass at age $0$;
 \item removing the Gaussian component accounting for adult mortality, setting $\pi_{1\,jt} =0$;
 	\item  characterizing the joint evolution of ${\boldsymbol{\tilde{\vartheta}}}_{jt}$ with a multivariate random walk, instead of independent innovations;
 	\item  characterizing the evolution of ${\boldsymbol{\tilde{\vartheta}}}_{jt}$ via random-walks with Student-$t$ innovations with $5$ degrees of freedom, instead of Gaussian distributions;
 \item replacing informative priors distributions in \cref{eq:prior} with flat uniform priors.
 \end{itemize}

All modifications are implemented and estimated with \texttt{nimble}, relying on the same settings as in \cref{sec:appeu}. Model evaluation is performed via marginal likelihood -- estimated via the harmonic mean estimator \citep[e.g.,][]{raftery:2006} -- and comparing the \textsc{mse} and \textsc{mae} of predictions obtained via posterior medians.
The empirical findings are reported in \cref{app:LL}, and suggest that the current implementation for \name{} is the best performing model in terms of marginal likelihood and other measures.
For example, modeling old-age mortality with a scaled Beta or characterizing infant mortality with an Half-normal distribution do not improve the fit of the current specification for \name{}; similar results are obtained with more complex specifications for the dynamic prior distribution or non-informative elicitation, confirming the importance of the proposed specifications.
Interestingly, current results indicate that the specification without adult mortality performs better for female than for male mortality, confirming that the inclusion of a separate component devoted to adult mortality is fundamental for modeling male mortality, but also provides concrete benefits for female mortality.

\begin{table}[tb]
\centering
\begin{tabular}{lccc|ccc}
&	\multicolumn{3}{c}{\textsc{males}} &	\multicolumn{3}{c}{\textsc{females}}\\
\cmidrule{2-7}
&$-\log p(\mathbf{D})$ &  \textsc{mae} & \textsc{mse} & $-\log p(\mathbf{D})$ &  \textsc{mae} & \textsc{mise} \\
 \toprule
\name{}                           & ${55.65}$& 0.10  & 10.19  & ${46.2}$ & ${0.08}$ & ${10.21}$ \\
	Beta (old-age mortality)          & $453.55$        & 0.13  & 33.83  & $307.53$        & 0.92            & 98.72 \\
	Half-normal (infant mortality)    & $176.03$        & 0.15  & 9.79   & $164.07$        & 0.95            & 98.72 \\
Removing adult mortality & $264.20$        & 0.88  & 30.79  & $88.23$         & 0.74            & 83.74 \\
	Multivariate random-walk          & $301.52$        & 8.78  & 116.76 & $205.52$        & 4.75            & 135.77\\
	Student-$t$ innovations           & $254.07$        & 0.90  & 10.26  & $178.23$        & 0.91            & 96.24 \\
	\name{}, uniform prior            & ${359.70}$      & 2.35 & 175.14 & ${466.52}$      & 2.01            & 121.61\\
 \bottomrule
\end{tabular}
\caption{Evaluation of model fit: $-\log p(\mathbf{D})$ refer to the logarithm of the marginal likelihood, changed of sign; \textsc{mse} and \textsc{mae} refer to Mean Squared Error and Mean Absloute Error of the posterior medians, respectively. Smaller values indicate better fit.}
\label{app:LL}
\end{table}

\section{Conclusion}

In this article we have introduced a dynamic model based on a Skewed distribution function (\name{}) for modeling and forecasting the age-at-death distribution of multiple countries across time.
The proposed method automatically regularize diverging scenarios, and allows to obtain predictions which are coherent across sub-populations.
In addition, this strategy leads to a borrowing of information which improved the quality of the fit and of the forecasts, compared with state-of-the-art methods for mortality forecasting.
Another important advantage of the proposed method is that it provides coherent forecasts without the need of choosing a reference population, which is instead naturally derived from data. 
The choice of reference population, in fact, is still an open issue of coherent forecast models and has received notable attention from scholars in recent years \citep{Booth:2020,kjaergaard:2016}.

We have focused directly on the age-distribution for the number of deaths, considering the total number of deaths as a given quantity and relying on a Multinomial likelihood.
This approach allows to focus directly on the distributions of deaths across different ages, accounting for the heterogeneity in the overall mortality trends.
For future developments, it might be useful to include a further dynamic level in the hierarchy and model the process of death counts $n_{j\,t}$.
Such an aim can be achieved, for example, using an hierarchical Poisson or Negative-Binomial dynamic model.

\section*{Acknowledgments}
This work was supported by the grant \textsc{miur–prin} 2017 project 20177BR-JXS.
The authors would like to thank David Dunson, Sally Paganin and Tommaso Rigon for their comments and the fruitful discussions.

\bibliographystyle{apalike}
\bibliography{99BIB.bib}
\end{document}


\maketitle
\section{Posterior Sampling}
In this section we illustrate the main steps to perform posterior inference under \textsc{dysm}. 
Following Section 2.1 of the paper, the model can be formulated as
\begin{equation}
	(\mathbf{D}_{jt} \mid n_{jt}, \mathbf{p}_{jt}) \sim \mbox{\textsc{multinom}}(n_{jt}, \mathbf{p}_{jt}), \quad j = 1, \dots, p, \quad t = 1, \dots, T.
	\label{eq:multi}
\end{equation}
with ${\mathbf{D}_{jt} = [D_{0\,jt}, \dots, D_{110\,jt}]}$ denoting the number of deaths and as $\mathbf{p}_{jt} = [p_{0\,jt}, \dots, p_{110\,jt}]$ the probabilities for the age-at-death distribution, where the elements $D_{x\,jt}$ denote the number of deaths at age $x$, in country $j$ during year $t$.
The vector of probabilities is induced discretizing a density function $f$ as
\begin{equation}
	\label{eq:int}
	p_{x\,jt} = \int_{x-1/2}^{x+1/2} f(z; \boldsymbol{\vartheta}_{jt}) \,\dif z = F(x+1/2; \boldsymbol{\vartheta}_{jt}) -  F(x-1/2; \boldsymbol{\vartheta}_{jt}), \quad x=0,\dots,110,
\end{equation}
which is specified as a mixture density function with
\begin{equation}
	\label{eq:mixf}
	f(x; \boldsymbol{\vartheta}_{jt}) =\pi_{0\,jt} \mathds{1}[x=0]  + 
	\pi_{1\,jt}\frac{1}{\sigma_{jt}}\phi\left(\frac{x - \mu_{jt}}{\sigma_{jt}}\right)  + \pi_{2j\,t} \frac{2}{\omega_{jt}}\phi \left(\frac{x - \xi_{jt}}{\omega_{jt}}\right) \Phi\left(\alpha_{jt}\frac{x - \xi_{jt}}{\omega_{jt}} \right).
\end{equation}
\noindent
For computational convenience, the model in parametrized in terms of uncosntrained parameters as 
\begin{equation}
	\widetilde{\boldsymbol{\vartheta}}_{jt} = (\tilde{\vartheta}_{jt \,1}, \dots,  \tilde{\vartheta}_{jt \,7})\coloneqq \left(\log\left(\frac{\pi_{1\,jt}}{1-\pi_{1\,jt}}\right), \log\left(\frac{\pi_{2\,jt}}{1-\pi_{1\,jt}-\pi_{2\,jt}}\right), \mu_{jt}, \log(\sigma_{jt}), \xi_{jt}, \log(\omega_{jt}), \alpha_{jt} \right),
	\label{eq:repar}
\end{equation}
and the prior distributions are specified as 
\begin{equation}
\begin{split}
	\tilde{\vartheta}_{j0 \, k} &\sim \mathcal{N}(m_{k}, s_{k}^2) \\
	(\tilde{\vartheta}_{jt \, k} \mid \tilde{\vartheta}_{j\,t-1\,k}, \beta_{j\,k}, \eta_{j\,k}^2) &\sim \mathcal{N}(  \beta_{j\,k} +  \tilde{\vartheta}_{j\,t-1\,k}, \eta_{j\,k}^2), \quad t = 1, \dots, T, \\
		\beta_{j\, k} \simI \mathcal{N} (m_{\beta_k}, s_{\beta_k}), &\quad \eta_{j\,k}^2 \simI \mbox{Inv-Gamma} (\mbox{a}_{k}, \mbox{b}_k) \quad \mbox{for} \,\, j = 1, \dots, p,
\end{split}
\label{eq:ts_mod}
\end{equation}
to induce an auto-regressive specification across time.

Posterior sampling is performed with the R package \texttt{nimble} \citep{nimblePack,nimbleArt}, which relies on an adaptive Metropolis-Hasting for non-conjugate step (updates of $\tilde{\vartheta}_{jt\,k}$) and a Gibbs Sampler for conjugate updates ($\beta_{jk}$ and $\eta^2_{jk}$).
The Metropolis-Hastings steps in \texttt{nimble} follow \citet{haario2001adaptive}, leveraging Gaussian random-walks with an adaptation of the variance of the proposal; refer to \citet{haario2001adaptive,nimblePack} for further details.

Specifically, the algorithm to conduct posterior inference under \textsc{dysm} follows the following steps. At each iteration $b$ in $1, \dots, B$, the Metropolis-within-Gibbs algorithm updates each multivariate parameter $\tilde{\vartheta}_{t\,k}=(\tilde{\vartheta}_{t\,1k}, \dots \tilde{\vartheta}_{t\,pk})$ recursively for $t=2,\dots,T$ as follows:
\begin{enumerate}
	\item Propose a candidate value $\tilde{\vartheta}_{t \, k}^\star$ from a Gaussian random-walk proposal $g(\tilde{\vartheta}_{t \, k}^\star, \tilde{\vartheta}_{t \, k}^{(b-1)})$ as in \citet{haario2001adaptive}.
	\item Accept the proposed value with probability $\min\{1, q(\tilde{\vartheta}_{t \, k}^\star)g(\tilde{\vartheta}_{t \, k}^\star, \tilde{\vartheta}_{t \, k}^{(b-1)}) / q(\tilde{\vartheta}_{t \, k}^{b-1}) (\tilde{\vartheta}_{t \, k}^{(b-1)}, \tilde{\vartheta}_{t \, k}^{\star})\}$, where $q(\tilde{\vartheta}_{t \, k})$ denotes the unnormalized posterior density evaluated at $\tilde{\vartheta}_{t \, k}$. 
		Considering the likelihood contribution for all countries, this quantity is proportional to the multinomial likelihood outlined in \cref{eq:multi}, multiplied by the density of the auto-regressive Gaussian priors, conditionally on the value at time $t-1$:
		\begin{equation}
			q(\tilde{\vartheta}_{t \, k})  \propto
			\prod_{j=0}^p\left[ \prod_{x=0}^{110} p_{x\,jt}^{D_{x\,jt}} \right] \phi(\tilde{\vartheta}_{jt \, k};  \beta_{j\,k}^{(b)} +  \tilde{\vartheta}_{j\,t-1\,k}^{(b)}, {\eta_{j\,k}^2}^{(b)}).
	\end{equation}
\item If the proposed value is accepted, set $\tilde{\vartheta}_{t\,k}^{(b)} =  \tilde{\vartheta}_{t\,k}^{\star}$; otherwise,  $\tilde{\vartheta}_{t\,k}^{(b)} =  \tilde{\vartheta}_{t\,k}^{(b-1)}$.
	\end{enumerate}
	Note that the likelihood function depends on the model parameters via $p_{x\,jt}  = p_{x\,jt}(\tilde{\vartheta}_{tj\,1},\dots, \tilde{\vartheta}_{tj\,7})$, computed according to \cref{eq:int}.
	Since $f(x;\cdot)$ in \cref{eq:int} is a mixture density function, its cumulative distribution function (\textsc{cdf}) $F(x, \cdot)$ corresponds to a mixture of \textsc{cdf}s, with same mixing weights.

	The parameters characterizing the random-walk components of \cref{eq:ts_mod} are updated considering the joint distribution for
$(\tilde{\vartheta}_{j1\,k}, \dots, \tilde{\vartheta}_{jT\,k})$, which correspond to a multivariate Gaussian with
$$(\tilde{\vartheta}_{j1\,k}, \dots, \tilde{\vartheta}_{jT\,k}) \sim \mathcal{N}(\beta_{j\,k}\mathds{1}_T, \eta^2_{j\,k}\Omega^{-1})$$
and $\Omega$ sparse $T\times T$ precision matrix with main diagonal equal to $(1, 2, \dots, 2, 1)$, first minor diagonals equal to $-1$ and remaining elements equal to $0$; see, for example, \citet[Section 9.2.1]{lindsey2004statistical} for a derivation of the precision matrix of an \textsc{ar(1)} process.
This simple expression allows to recast the problem into conjugate Gaussian updates for $\beta_{j\,k}$ and into conjugate Inverse-Gamma updates for $\eta^2_{j\,k}$, with full conditional distributions equal to
\begin{equs}
	(\beta_{j\,k} \mid -) &\sim \mathcal{N}\left( (\mathds{1}^\intercal \Omega \mathds{1}/ \eta^2_{j\,k} + 1/s_{\beta_k})^{-1} (\mathds{1}^\intercal \Omega \boldsymbol{\tilde{\vartheta}}_{j\,k} /\eta^2_{j\,k} + m_{\beta_k}/s_{\beta_k}) , (\mathds{1}^\intercal \Omega \mathds{1}/\eta^2_{j\,k} + 1/s_{\beta_k})^{-1} \right),\\
	(\eta^2_{j\,mk} \mid - ) &\sim \mbox{Inv-Gamma}(a_k + T/2, b_k + \mathds{1}^\intercal (L \boldsymbol{\tilde{\vartheta}}_{j\,k} - \beta_{j\,k}\mathds{1}_T)/2 ),
\end{equs}
where $L$ is such that $L^\intercal L = \Omega^{-1}$.

\section{Additional Simulations}
In this section, we expand the results reported in Section 4 of the paper, illustrating the country-specific performance of \textsc{dysm} against the competitors. In-sample performances are reported in \cref{app:ins_full1,app:ins_full2}, while out-of-sample metrics are reported in \cref{app:oos_full1,app:oos_full2}.
Considering, for example, in-sample performance reported in \cref{app:ins_full1}-\ref{app:ins_full2}, we observe that \name{} performs slightly worse than some competitors only in the female Danish and Dutch populations. 
These results might be due to the specific structures of such populations, which deviate from the overall patterns and might be subject to an excessive amount of shrinkage by \textsc{dysm}; refer also to \citep{danish:20,lindahl:2016} for details on the Danish case.
When out-of-sample performance is considered, only results for the male Spanish and Swedish populations indicate that \name{} is slightly less accurate than some competitors in these countries.
These results might be due to some specific characteristics of these countries, which notably affected mortality and life expectancy and are notably different from the overall trend, resulting in worse performances when coherency across predictions is imposed.

\include{iistables}
\include{oostables}

\section{Alternative parametric specifications}
Finally, in this section we illustrate in details the approaches used in Section 3.4 of the paper.
The competing approaches rely on the same Multinomial specification outlined in \cref{eq:multi,eq:int}, while the underlying probability density function in \cref{eq:mixf} and the prior distributions in \cref{eq:ts_mod} are specified differently.
Using a more general notation, we generalize \cref{eq:mixf} as
\begin{equation}
	\label{threemix}
	f(x; \boldsymbol{\vartheta}_{jt}) =\pi_{0\,jt} g_0(x;   \vartheta_{0\,jt})+ 
	\pi_{1\,jt}g_1(x;   \vartheta_{1\,jt})+  \pi_{2j\,t} g_2(x;   \vartheta_{2\,jt}),
\end{equation}
where $g_0,g_1$ and $g_2$ denote the probability density function characterizing the mixure components.
\begin{itemize}
	\item The first competing approach relies on the same specification as \textsc{dysm}, but replaces $g_2$ with the density of a rescaled Beta-Distribution in the interval $75-100$. This choiche leads to 
$$g_2(x) = \dfrac{\Gamma(a_{jt} + b_{jt})}{\Gamma(a_{jt})\Gamma(b_{jt})}\left(\dfrac{x-75}{110-75}\right)^{a_{jt}}\left(1-\dfrac{x-75}{110-75}\right)^{b_{jt}}$$
\item The second specification relies on an half-normal distribution for infant mortality. This leads to 
$$g_1(x) ={\frac  {{\sqrt  {2}}}{\gamma_{jt} {\sqrt  {\pi }}}}\exp \left(-{\frac  {x^{2}}{2\gamma_{jt} ^{2}}}\right) $$
\item The third specification removes the component devoted to adult mortality. This setting can be obtained simply setting $\pi_{1\,jt}=0$.
\end{itemize}
The remaining competitors are obtained with modifications on the prior distribution of \cref{eq:ts_mod}. Student-$t$ innovations are obtained replacing the Gaussian distributions in \cref{eq:ts_mod} with Student-$t$ with $\nu=5$ degrees of freedom. Similarly, multivariate random walks can be obtained considering the vector $\tilde{\boldsymbol{\vartheta}}_{jt} \sim \mathcal{N}_7(\tilde{\boldsymbol{\vartheta}}_{j\,t-1}, \boldsymbol{\Sigma})$, with $\boldsymbol{\Sigma}$ full $7\times7$ covariance matrix assigned to an Inverse-Wishart prior.

\printbibliography

%% file: pack.tex
\usepackage[T1]{fontenc}
\usepackage{amsfonts,mathtools, amsmath,amsthm,amssymb,dsfont,mhequ}
\usepackage{graphicx,booktabs,xcolor, array,multirow}
\usepackage{adjustbox}
\usepackage{pdflscape}
\usepackage[top = 2cm , bottom= 2cm , left= 2cm , right= 2cm]{geometry}
\usepackage[colorlinks,citecolor=rr, linkcolor=rr]{hyperref}
\usepackage[labelfont=bf, footnotesize]{caption}
\definecolor{prp}{HTML}{6C71C4}
\definecolor{bl}{HTML}{0076A1}
\definecolor{gg}{HTML}{6D6D6D}
\definecolor{rr}{HTML}{f92672}

\usepackage{appendix,float}
\usepackage{authblk}
\usepackage{setspace}
\usepackage[shortlabels]{enumitem}

\usepackage{rotating}

\usepackage{mdframed}
\mdfdefinestyle{st}{%
    linecolor=bl,
   linewidth=2pt,
        }

\usepackage{titlesec}
\titleformat{\section}
  {\centering\scshape}
  {\thesection}
  {.5em}
  {\MakeUppercase}

\titleformat{\subsection}
  {\normalfont\scshape\itshape}{\thesubsection}{1em}{}

\usepackage[ruled,norelsize,vlined]{algorithm2e}
\usepackage[noabbrev,capitalise]{cleveref}



%% file: dag.tex

\begin{figure}[bt]
	\centering
	\begin{tikzpicture}

	\tikzstyle{main}=[minimum size = 14mm, thick, draw =black!80, node distance = 5mm]
	\tikzstyle{connect}=[-latex, thick]
	
	\node[main, circle, fill = white!100] (theta1) {$\btheta_{j1}$ };
	\node[main, circle, fill = white!100] (theta2) [right=of theta1]{$\btheta_{j2}$ };
	\node[main, circle, fill = white!100,draw=white] (dots1) [right=of theta2] {$\cdots$ };
	\node[main, circle, fill = white!100] (thetat) [right=of dots1] {$\btheta_{jt}$ };
	\node[main, circle, fill = white!100,draw=white] (dots2) [right=of thetat] {$\cdots$ };
	\node[main, circle, fill = white!100] (thetan1) [right=of dots2] {$\btheta_{j_{\,T-1}}$ };
	\node[main, circle, fill = white!100] (thetan) [right=of thetan1] {$\btheta_{j_{\,T} }$ };

	\node[main, fill = white!10] (z1) [below=of theta1] {$\mathbf{p}_{j1}$ };
	\node[main, fill = white!10] (z2) [below=of theta2] {$\mathbf{p}_{j2}$ };
	\node[main, fill = white!10,draw=white] (dots11) [below=of dots1] {$\cdots$ };
	\node[main, fill = white!10] (zt) [below=of thetat] {$\mathbf{p}_{jt}$ };
	\node[main, fill = white!10,draw=white] (dots12) [below=of dots2] {$\cdots$ };
	\node[main, fill = white!10] (zn1) [below=of thetan1]{$\mathbf{p}_{j_{\,T-1}}$ };
	\node[main, fill = white!10] (zn) [below=of thetan]{$\mathbf{p}_{j_{\,T}}$ };
	
\node[main, fill = black!10] (y1) [below=of z1] {$\mathbf{D}_{j1}$};
	\node[main, fill = black!10] (y2) [below=of z2] {$\mathbf{D}_{j2}$};
	\node[main, fill = white!10,draw=white] (dots21) [below=of dots11] {$\cdots$ };
\node[main, fill = black!10] (yt) [below=of zt] {$\mathbf{D}_{jt}$};
	\node[main, fill = white!10,draw=white] (dots22) [below=of dots12] {$\cdots$ };
	\node[main, fill = black!10] (yn1) [below=of zn1]{$\mathbf{D}_{j_{\,T-1}}$};
	\node[main, fill = black!10] (yn) [below=of zn] {$\mathbf{D}_{j_{\,T}}$};
	\node[rectangle, inner sep=6.5mm, draw=black!100, fit = (theta1) (yn)] (rr) {};
	\node[] at (11,-5) {{\small{for country $j=1, \ldots, p$}}};
	
	\path        (theta1) edge [connect] (z1);
	\path        (theta2) edge [connect] (z2);
	\path    (thetat) edge [connect] (zt);
	\path        (thetan1) edge [connect] (zn1);
	\path      (thetan) edge [connect] (zn);
	\path        (z1) edge [connect] (y1);
	\path        (z2) edge [connect] (y2);
	\path    (zt) edge [connect] (yt);
	\path      (zn1) edge [connect] (yn1);
	\path      (zn) edge [connect] (yn);
	\path        (theta1) edge [connect] (theta2);
	\path        (theta2) edge [connect] (dots1);
	\path        (dots1) edge [connect] (thetat);
	\path        (thetat) edge [connect] (dots2);
	\path        (dots2) edge [connect] (thetan1);
	\path        (thetan1) edge [connect] (thetan);

	\end{tikzpicture}
	\caption{Graphical illustration of the dynamic component of \name{}. Solid circles, white squares and gray squares denote vectors of dynamic parameters, probabilities characterizing the age-at-death distribution and observed number of deaths, respectively.}
	\label{dagM}
\end{figure}

%% file: dag2.tex
\begin{figure}[t]
	\centering
	\begin{tikzpicture}
	\tikzstyle{main}=[minimum size = 15mm, thick, draw =black!80, node distance = 7mm, every node/.style={anchor=base,text depth=.5ex,text height=2ex,text width=1em}]
	\tikzstyle{connect}=[-latex, thick]
	
	\node[main, fill = white!100] (j1) {$\beta_{1\,k}, \eta_{1\,k},  \left(\tilde{\vartheta}_{1t \, k}\right)_{t=0}^T$ };
	\node[main, fill = white!100,draw=white] (d1) [right=of j1] {$\cdots$ };
	\node[main, fill = white!100] (jn) [right=of d1]  {$\beta_{p\,k}, \eta_{p\,k},  \left(\tilde{\vartheta}_{pt \, k}\right)_{t=0}^T$ };
	\node[main, fill = white!100] (prior) [above=of d1] {$m_k, s_k, m_{\beta_k}, s_{\beta_k}, a_k, b_k $};
	\node[main, fill = white!100, draw = white] (empthy)[above=of prior] {};
	\node[rectangle, inner sep=6.5mm, draw=black!100, fit = (prior) (j1) (jn)] (rr) {};
	\node[] at (7.5,-1.2) {{\small{$k=1, \ldots, 7$}}};
	\path        (prior) edge [connect, dashed] (j1);
	\path        (prior) edge [connect, dashed] (jn);
	
	\end{tikzpicture}
	\caption{Graphical illustration of the hierarchical component of \name{}, inducing borrowing of information and coherent predictions across countries, with $j=1, \dots,p$. Top level block denotes fixed hyper-parameters, while bottom blocks denote model's parameters.} 
	\label{dagM2}
\end{figure}

%% file: iistables.tex
\begin{sidewaystable}
	\begin{tabular}{rllccccccc}
 & & & \textsc{aus} & \textsc{che} & \textsc{deu} &\textsc{dnk} & \textsc{esp} & \textsc{fin}\\
\toprule
	\parbox[t]{2mm}{\multirow{5}{*}{\rotatebox[origin=c]{90}{\textsc{females}}}} 
& \textsc{mae} & \name{}       & 1                   & 1                   & 1                   & 1                   & 1                   & 1                   \\
&              & Hyndman-Ullah & 1.52 $[1.44, 1.66]$ & 1.31 $[1.23, 1.49]$ & 1.59 $[1.46, 1.66]$ & 0.93 $[0.83, 1.16]$ & 1.74 $[1.61, 1.82]$ & 1.63 $[1.54, 1.69]$ \\
&              & Lee-Carter    & 1.52 $[1.44, 1.65]$ & 1.34 $[1.25, 1.49]$ & 1.58 $[1.36, 1.64]$ & 0.93 $[0.83, 1.16]$ & 1.72 $[1.59, 1.80]$ & 1.62 $[1.55, 1.71]$ \\
&              & Li-Lee        & 1.46 $[1.36, 1.56]$ & 1.22 $[1.16, 1.42]$ & 1.27 $[0.91, 1.61]$ & 0.89 $[0.78, 1.14]$ & 0.94 $[0.61, 1.44]$ & 1.57 $[1.46, 1.66]$ \\
&              & \textsc{mem5} & 1.21 $[1.20, 1.22]$ & 1.04 $[1.04, 1.05]$ & 1.21 $[1.18, 1.28]$ & 1.02 $[0.95, 1.11]$ & 1.45 $[1.39, 1.52]$ & 1.30 $[1.29, 1.33]$ \\
&              & Oeppen        & 1.31 $[1.21, 1.41]$ & 1.11 $[1.01, 1.24]$ & 0.84 $[0.76, 1.25]$ & 0.82 $[0.76, 0.91]$ & 1.46 $[1.36, 1.69]$ & 1.18 $[1.09, 1.43]$ \\
\cmidrule{2-9}                                                                                                                                                                                                
& \textsc{mse} & \name{}       & 1                   & 1                   & 1                   & 1                   & 1                   & 1                   \\
&              & Hyndman-Ullah & 2.64 $[2.42, 3.29]$ & 2.45 $[2.16, 2.96]$ & 3.51 $[3.14, 4.07]$ & 1.05 $[0.72, 1.76]$    & 4.64 $[3.89, 5.31]$ & 3.54 $[3.02, 4.00]$ \\
&              & Lee-Carter    & 2.65 $[2.42, 3.27]$ & 2.55 $[2.22, 2.99]$ & 3.48 $[2.95, 4.03]$ & 0.99 $[0.73, 1.78]$ & 4.56 $[3.94, 5.28]$ & 3.57 $[3.04, 4.00]$ \\
&              & Li-Lee        & 2.56 $[2.15, 3.02]$ & 2.18 $[1.92, 2.76]$ & 2.73 $[1.56, 3.98]$ & 0.94 $[0.67, 1.71]$ & 1.47 $[0.54, 3.27]$ & 3.39 $[2.77, 3.85]$ \\
&              & \textsc{mem5} & 1.74 $[1.69, 1.85]$ & 1.45 $[1.43, 1.52]$ & 2.19 $[1.87, 2.91]$ & 1.19 $[1.03, 1.48]$ & 3.52 $[3.07, 3.92]$ & 2.28 $[2.13, 2.35]$ \\
&              & Oeppen        & 2.08 $[1.76, 2.45]$ & 1.70 $[1.44, 2.13]$ & 1.38 $[1.10, 2.25]$ & 0.95 $[0.73, 1.11]$ & 3.45 $[3.04, 4.53]$ & 1.81 $[1.56, 2.85]$ \\
\midrule
\parbox[t]{2mm}{\multirow{5}{*}{\rotatebox[origin=c]{90}{\textsc{males}}}} 
& \textsc{mae} & \name{}       & 1                   & 1                   & 1                   & 1                   & 1                   & 1                   \\
&              & Hyndman-Ullah & 2.50 $[1.84, 2.69]$ & 1.73 $[1.43, 2.00]$ & 2.59 $[1.88, 2.74]$ & 1.18 $[0.93, 1.39]$    & 2.28 $[2.10, 2.34]$ & 1.76 $[1.68, 2.08]$ \\
&              & Lee-Carter    & 2.50 $[1.83, 2.71]$ & 1.72 $[1.47, 2.02]$ & 2.58 $[1.87, 2.73]$ & 1.17 $[0.93, 1.39]$    & 2.27 $[2.21, 2.30]$ & 1.77 $[1.69, 2.10]$ \\
&              & Li-Lee        & 2.36 $[1.81, 2.69]$ & 1.61 $[1.38, 1.84]$ & 2.47 $[1.86, 2.69]$ & 0.97 $[0.91, 1.07]$ & 1.79 $[1.57, 2.00]$ & 1.68 $[1.62, 1.84]$ \\
&              & \textsc{mem5} & 2.00 $[1.87, 2.09]$ & 1.72 $[1.69, 1.74]$ & 2.15 $[2.05, 2.22]$ & 1.53 $[1.50, 1.56]$ & 2.28 $[2.22, 2.32]$ & 1.48 $[1.46, 1.51]$ \\
&              & Oeppen        & 2.31 $[1.89, 2.71]$ & 1.44 $[1.35, 1.83]$ & 1.32 $[1.10, 2.48]$ & 0.98 $[0.92, 1.32]$ & 1.76 $[1.33, 2.27]$ & 1.58 $[1.45, 1.71]$ \\
\cmidrule{2-9}                                                                                                                                                                                                
& \textsc{mse} & \name{}       & 1                    & 1                   & 1                   & 1                   & 1                    & 1                   \\
&              & Hyndman-Ullah & 8.01 $[4.22, 10.32]$ & 3.85 $[2.99, 5.28]$ & 7.08 $[5.69, 7.56]$ & 1.26 $[1.04, 2.27]$ & 7.23 $[6.53, 8.43]$  & 3.40 $[3.14, 5.34]$ \\
&              & Lee-Carter    & 8.00 $[4.23, 10.27]$ & 3.85 $[3.08, 5.39]$ & 7.01 $[5.67, 7.58]$ & 1.26 $[1.04, 2.27]$ & 7.23 $[6.71, 8.49]$  & 3.61 $[3.14, 5.33]$ \\
&              & Li-Lee        & 7.36 $[4.18, 10.15]$ & 3.60 $[2.96, 4.60]$ & 6.92 $[5.56, 7.53]$ & 1.17 $[0.96, 1.41]$ & 5.56 $[4.50, 6.12]$  & 3.32 $[2.98, 4.44]$ \\
&              & \textsc{mem5} & 4.71 $[4.57, 4.80]$  & 3.41 $[2.99, 3.87]$ & 6.01 $[3.48, 9.17]$ & 2.78 $[2.51, 2.98]$ & 7.92 $[6.59, 10.47]$ & 2.47 $[2.38, 2.72]$ \\
&              & Oeppen        & 6.63 $[4.30, 10.77]$ & 3.05 $[2.62, 4.30]$ & 2.75 $[2.13, 5.58]$ & 1.17 $[1.02, 2.13]$ & 5.18 $[3.62, 6.91]$  & 3.07 $[2.48, 3.72]$ \\
\bottomrule
\end{tabular}
\caption{In-sample relative performance. Median across rolling windows, stratified by country (\textsc{aus}, \textsc{che}, \textsc{deu},\textsc{dnk}, \textsc{esp}, \textsc{fin}). First and third quartiles in squared brackets.}
\label{app:ins_full1}
\end{sidewaystable}

\begin{sidewaystable}
	\begin{tabular}{rllccccccc}
 & & & \textsc{fra} & \textsc{gbr} & \textsc{ita} & \textsc{nld} & \textsc{nor} & \textsc{swe} \\ 
\toprule
	\parbox[t]{2mm}{\multirow{5}{*}{\rotatebox[origin=c]{90}{\textsc{females}}}} 
& \textsc{mae} & \name{}       & 1                   & 1                   & 1                   & 1                   & 1                   & 1                \\
&              & Hyndman-Ullah & 1.47 $[1.39, 1.72]$ & 1.48 $[1.04, 1.86]$ & 1.72 $[1.42, 1.92]$ & 0.97 $[0.71, 1.29]$ & 1.06 $[0.99, 1.17]$ & 1.34 $[1.26, 1.38]$ \\
&              & Lee-Carter    & 1.44 $[1.33, 1.68]$ & 1.47 $[1.03, 1.86]$ & 1.64 $[1.40, 1.90]$ & 0.97 $[0.70, 1.28]$ & 1.05 $[0.99, 1.16]$ & 1.33 $[1.25, 1.37]$ \\
&              & Li-Lee        & 1.31 $[0.46, 1.47]$ & 0.82 $[0.76, 1.12]$ & 1.48 $[1.32, 1.81]$ & 0.95 $[0.67, 1.18]$ & 1.04 $[0.98, 1.11]$ & 1.24 $[1.19, 1.31]$ \\
&              & \textsc{mem5} & 1.26 $[1.23, 1.29]$ & 1.32 $[1.22, 1.56]$ & 1.29 $[1.26, 1.35]$ & 1.02 $[1.01, 1.04]$ & 1.07 $[1.05, 1.08]$ & 1.24 $[1.14, 1.29]$ \\
&              & Oeppen        & 1.17 $[1.10, 1.32]$ & 1.45 $[1.04, 1.83]$ & 1.25 $[1.12, 1.38]$ & 0.72 $[0.64, 0.90]$ & 0.95 $[0.90, 1.11]$ & 1.12 $[1.09, 1.18]$ \\
\cmidrule{2-9}                                                                                                                                                                                                
& \textsc{mse} & \name{}       & 1                & 1                & 1                & 1                & 1                & 1                \\
&              & Hyndman-Ullah & 3.65 $[3.13, 5.13]$ & 2.44 $[1.39, 3.75]$ & 4.76 $[3.03, 5.64]$ & 1.46 $[0.70, 2.66]$ & 1.55 $[1.27, 1.93]$ & 2.56 $[2.05, 2.69]$ \\
&              & Lee-Carter    & 3.53 $[3.10, 5.08]$ & 2.44 $[1.39, 3.75]$ & 4.66 $[3.01, 5.60]$ & 1.47 $[0.69, 2.64]$ & 1.55 $[1.28, 1.92]$ & 2.46 $[2.04, 2.66]$ \\
&              & Li-Lee        & 3.04 $[0.35, 3.91]$ & 1.03 $[0.70, 1.63]$ & 3.51 $[2.80, 5.13]$ & 1.44 $[0.61, 2.20]$ & 1.43 $[1.24, 1.77]$ & 2.20 $[2.00, 2.54]$ \\
&              & \textsc{mem5} & 2.69 $[2.53, 3.05]$ & 2.26 $[2.08, 2.65]$ & 2.68 $[2.40, 3.34]$ & 1.48 $[1.39, 1.64]$ & 1.51 $[1.40, 1.62]$ & 2.01 $[1.65, 2.35]$ \\
&              & Oeppen        & 2.51 $[1.88, 3.03]$ & 2.38 $[1.40, 3.54]$ & 2.63 $[2.11, 3.21]$ & 0.73 $[0.59, 1.28]$ & 1.33 $[1.11, 1.77]$ & 1.88 $[1.65, 2.14]$ \\
\midrule
\parbox[t]{2mm}{\multirow{5}{*}{\rotatebox[origin=c]{90}{\textsc{males}}}} 
& \textsc{mae} & \name{}       & 1                & 1                 & 1                  & 1                & 1                & 1                 \\
&              & Hyndman-Ullah & 2.15 $[1.73, 2.33]$ & 2.76 $[1.92, 3.20]$  & 2.86 $[1.89, 3.68]$   & 1.55 $[0.97, 2.07]$ & 1.17 $[0.93, 1.67]$ & 1.98 $[1.27, 2.39]$  \\
&              & Lee-Carter    & 2.08 $[1.70, 2.26]$ & 2.77 $[1.91, 3.21]$  & 2.86 $[1.88, 3.62]$   & 1.57 $[0.94, 2.06]$ & 1.17 $[0.93, 1.67]$ & 1.97 $[1.27, 2.35]$  \\
&              & Li-Lee        & 1.88 $[1.64, 2.07]$ & 2.72 $[1.84, 3.14]$  & 2.60 $[1.84, 3.02]$   & 1.52 $[0.93, 2.02]$ & 1.13 $[0.91, 1.63]$ & 1.78 $[1.25, 2.29]$  \\
&              & \textsc{mem5} & 1.83 $[1.79, 1.97]$ & 2.26 $[2.16, 2.33]$  & 2.75 $[2.59, 2.85]$   & 1.81 $[1.58, 1.93]$ & 1.52 $[1.51, 1.55]$ & 1.88 $[1.83, 1.93]$  \\
&              & Oeppen        & 1.54 $[1.35, 1.61]$ & 2.71 $[1.71, 3.10]$  & 1.83 $[1.60, 2.82]$   & 1.53 $[0.75, 2.04]$ & 1.16 $[0.96, 1.71]$ & 1.68 $[1.28, 2.36]$  \\
\cmidrule{2-9}                                                                                                                                                                                                
& \textsc{mse} & \name{}       & 1                & 1                 & 1                  & 1                & 1                & 1                 \\
&              & Hyndman-Ullah & 7.44 $[5.55, 8.29]$ & 9.53 $[5.06, 11.22]$ & 12.67 $[7.88, 18.54]$ & 3.02 $[1.15, 5.64]$ & 1.66 $[1.12, 3.42]$ & 4.67 $[1.93, 6.83]$  \\
&              & Lee-Carter    & 7.15 $[5.45, 8.06]$ & 9.51 $[5.02, 11.18]$ & 13.39 $[7.83, 18.45]$ & 3.07 $[1.11, 5.62]$ & 1.66 $[1.12, 3.44]$ & 4.64 $[1.93, 6.69]$  \\
&              & Li-Lee        & 6.12 $[5.09, 6.85]$ & 9.28 $[4.89, 11.06]$ & 10.57 $[7.31, 14.15]$ & 2.87 $[1.09, 5.39]$ & 1.59 $[1.05, 3.30]$ & 3.85 $[1.87, 6.52]$  \\
&              & \textsc{mem5} & 4.82 $[3.84, 7.52]$ & 7.34 $[5.01, 8.48]$  & 12.84 $[9.06, 15.53]$ & 3.90 $[3.56, 4.31]$ & 2.70 $[2.51, 3.20]$ & 3.99 $[3.52, 4.63]$  \\
&              & Oeppen        & 3.78 $[3.51, 4.57]$ & 9.34 $[4.05, 10.94]$ & 6.95 $[5.24, 12.84]$  & 2.84 $[0.79, 5.55]$ & 1.63 $[1.15, 3.55]$ & 3.38 $[1.95, 6.71]$  \\
\bottomrule
\end{tabular}
\caption{In-sample relative performance. Median across rolling windows, stratified by country (\textsc{fra}, \textsc{gbr}, \textsc{ita}, \textsc{nld}, \textsc{nor}, \textsc{swe}). First and third quartiles in squared brackets.}
\label{app:ins_full2}
	\end{sidewaystable}

%% file: oostables.tex
\begin{sidewaystable}
	\begin{tabular}{rllcccccc}
& & & \textsc{aus} & \textsc{che} & \textsc{deu} &\textsc{dnk} & \textsc{esp} & \textsc{fin} \\
\toprule
\parbox[t]{2mm}{\multirow{5}{*}{\rotatebox[origin=c]{90}{\textsc{females}}}}
& \textsc{mae} & \name{}       & 1                   & 1                   & 1                   & 1                   & 1                   & 1                   \\
&              & Hyndman-Ullah & 2.05 $[1.89, 2.16]$ & 1.66 $[1.46, 2.15]$ & 1.33 $[1.26, 1.42]$ & 1.08 $[0.88, 1.33]$ & 1.23 $[1.14, 1.59]$ & 1.09 $[0.91, 1.21]$ \\
&              & Lee-Carter    & 1.98 $[1.72, 2.19]$ & 1.62 $[1.37, 2.10]$ & 1.29 $[1.11, 1.40]$ & 1.07 $[0.81, 1.35]$ & 1.11 $[0.97, 1.38]$ & 1.04 $[0.84, 1.26]$ \\
&              & Li-Lee        & 1.97 $[1.72, 2.19]$ & 1.62 $[1.36, 2.10]$ & 1.29 $[0.78, 1.39]$ & 1.07 $[0.81, 1.35]$ & 0.82 $[0.54, 1.35]$ & 1.04 $[0.84, 1.26]$ \\
&              & \textsc{mem5} & 3.55 $[3.37, 3.86]$ & 3.06 $[2.65, 3.90]$ & 2.30 $[2.20, 2.45]$ & 1.72 $[1.19, 2.25]$ & 2.34 $[2.02, 2.95]$ & 1.95 $[1.63, 2.48]$ \\
&              & Oeppen        & 1.89 $[1.63, 2.83]$ & 2.19 $[1.42, 3.25]$ & 0.82 $[0.60, 1.11]$ & 1.25 $[0.91, 3.70]$ & 1.05 $[0.89, 1.30]$ & 1.28 $[0.93, 2.00]$ \\
\cmidrule{2-9}
& \textsc{mse} & \name{}       & 1                     & 1                     & 1                   & 1                     & 1                      & 1                    \\
&              & Hyndman-Ullah & 3.69 $[3.15, 4.50]$   & 2.57 $[2.01, 4.27]$   & 1.86 $[1.51, 2.12]$ & 1.10 $[0.84, 2.00]$   & 1.80 $[1.52, 3.04]$    & 1.20 $[0.87, 1.54]$  \\
&              & Lee-Carter    & 3.48 $[2.63, 4.28]$   & 2.51 $[1.67, 4.27]$   & 1.60 $[1.15, 2.05]$ & 1.14 $[0.68, 1.87]$   & 1.48 $[1.11, 2.16]$    & 1.10 $[0.75, 1.67]$  \\
&              & Li-Lee        & 3.47 $[2.62, 4.28]$   & 2.50 $[1.67, 4.27]$   & 1.59 $[0.62, 2.05]$ & 1.14 $[0.68, 1.87]$   & { 0.87} $[0.39, 2.07]$ & 1.10 $[0.75, 1.67]$  \\
&              & \textsc{mem5} & 10.45 $[8.76, 11.88]$ & 8.09 $[5.92, 13.37]$  & 4.96 $[4.31, 5.81]$ & 2.95 $[1.16, 5.21]$   & 6.17 $[4.78, 9.59]$    & 3.60 $[2.50, 5.78]$  \\
&              & Oeppen        & 3.45 $[2.44, 92.54]$  & 6.31 $[2.38, 150.17]$ & 0.94 $[0.50, 6.41]$ & 1.91 $[0.98, 197.90]$ & 1.55 $[1.18, 3.74]$    & 6.24 $[1.50, 49.16]$ \\
\midrule
\parbox[t]{2mm}{\multirow{5}{*}{\rotatebox[origin=c]{90}{\textsc{males}}}} 
& \textsc{mae} & \name{}       & 1                   & 1                   & 1                   & 1                   & 1                      & 1                   \\
&              & Hyndman-Ullah & 3.22 $[2.66, 3.56]$ & 2.27 $[2.02, 2.38]$ & 2.94 $[2.67, 3.05]$ & 1.15 $[0.72, 2.06]$ & 0.94 $[0.86, 1.18]$    & 1.47 $[1.27, 1.84]$ \\
&              & Lee-Carter    & 2.99 $[2.15, 3.45]$ & 2.19 $[1.95, 2.37]$ & 2.79 $[2.35, 3.03]$ & 1.14 $[0.74, 2.03]$ & 0.97 $[0.73, 1.14]$    & 1.44 $[1.15, 1.82]$ \\
&              & Li-Lee        & 2.94 $[2.15, 3.45]$ & 2.19 $[1.95, 2.36]$ & 2.79 $[2.35, 3.03]$ & 1.14 $[0.74, 2.02]$ & { 0.97} $[0.72, 1.14]$ & 1.44 $[1.15, 1.82]$ \\
&              & \textsc{mem5} & 5.39 $[4.83, 6.10]$ & 3.58 $[3.31, 3.88]$ & 4.51 $[3.81, 4.92]$ & 1.32 $[1.12, 2.73]$ & 1.67 $[1.53, 1.77]$    & 2.40 $[1.96, 3.02]$ \\
&              & Oeppen        & 3.10 $[2.50, 3.68]$ & 2.15 $[1.99, 2.53]$ & 1.96 $[1.51, 2.23]$ & 1.47 $[0.81, 2.10]$ & 0.96 $[0.50, 1.14]$    & 1.65 $[1.37, 2.64]$ \\
\cmidrule{2-9}
& \textsc{mse} & \name{}       & 1                      & 1                     & 1                      & 1                   & 1                   & 1                     \\
&              & Hyndman-Ullah & 10.26 $[6.83, 12.63]$  & 5.01 $[4.17, 5.43]$   & 9.02 $[6.86, 10.66]$   & 1.43 $[0.68, 4.22]$ & 1.03 $[0.86, 1.56]$ & 2.04 $[1.61, 3.32]$   \\
&              & Lee-Carter    & 9.12 $[4.54, 11.37]$   & 4.89 $[3.86, 5.29]$   & 7.84 $[6.01, 10.28]$   & 1.42 $[0.68, 4.03]$ & 1.11 $[0.60, 1.51]$ & 2.02 $[1.35, 3.29]$   \\
&              & Li-Lee        & 8.82 $[4.54, 11.37]$   & 4.89 $[3.86, 5.29]$   & 7.84 $[6.00, 10.28]$   & 1.42 $[0.68, 4.00]$ & 1.09 $[0.59, 1.51]$ & 2.02 $[1.35, 3.29]$   \\
&              & \textsc{mem5} & 25.68 $[20.59, 35.34]$ & 12.00 $[9.98, 13.97]$ & 19.54 $[15.15, 24.84]$ & 2.79 $[1.75, 6.42]$ & 3.27 $[2.77, 3.79]$ & 4.84 $[3.67, 8.29]$   \\
&              & Oeppen        & 11.68 $[7.72, 16.26]$  & 5.81 $[4.37, 24.00]$  & 5.47 $[3.98, 9.44]$    & 3.32 $[1.20, 4.90]$ & 1.16 $[0.35, 1.55]$ & 3.40 $[1.98, 111.99]$ \\
\bottomrule
	\end{tabular}
	\caption{Out-of-sample relative performance. Median across rolling windows, stratified by country (\textsc{aus}, \textsc{che}, \textsc{deu},\textsc{dnk}, \textsc{esp}, \textsc{fin}). First and third quartiles in squared brackets.}
	\label{app:oos_full1}
\end{sidewaystable}

\begin{sidewaystable}
	\begin{tabular}{rllccccccc}
  & &  & \textsc{fra} & \textsc{gbr} & \textsc{ita} & \textsc{nld} & \textsc{nor} & \textsc{swe} \\ 
  \toprule
  \parbox[t]{2mm}{\multirow{5}{*}{\rotatebox[origin=c]{90}{\textsc{females}}}}
  & \textsc{mae} & \name{}        & 1                   & 1                   & 1                   & 1                   & 1                   & 1 \\
  &              & Hyndman-Ullah  & 1.60 $[1.43, 1.71]$ & 1.47 $[1.12, 1.65]$ & 1.59 $[1.26, 1.88]$ & 1.13 $[0.70, 1.99]$ & 1.41 $[1.32, 1.54]$ & 1.15 $[1.09, 1.22]$  \\
  &              & Lee-Carter     & 1.57 $[1.24, 1.71]$ & 1.48 $[1.07, 1.60]$ & 1.39 $[1.20, 1.78]$ & 1.04 $[0.63, 1.94]$ & 1.43 $[1.31, 1.54]$ & 1.14 $[0.99, 1.25]$  \\
  &              & Li-Lee         & 1.41 $[0.69, 1.71]$ & 1.08 $[0.95, 1.34]$ & 1.39 $[1.20, 1.78]$ & 1.05 $[0.63, 1.94]$ & 1.43 $[1.31, 1.54]$ & 1.14 $[0.99, 1.25]$  \\
  &              & \textsc{mem5}  & 2.83 $[2.52, 3.00]$ & 2.29 $[1.85, 2.64]$ & 2.69 $[2.27, 3.17]$ & 1.86 $[1.24, 4.07]$ & 2.40 $[2.10, 2.76]$ & 2.08 $[1.91, 2.27]$  \\
  \cmidrule{2-9}                                                                                                                                                                                                
  & \textsc{mse} & \name{}       & 1                   & 1                   & 1                   & 1                    & 1                    & 1   \\
  &              & Hyndman-Ullah & 2.66 $[2.15, 3.09]$ & 1.99 $[1.28, 2.63]$ & 2.63 $[1.72, 3.51]$ & 1.42 $[0.68, 4.46]$  & 2.06 $[1.80, 2.48]$  & 1.43 $[1.35, 1.64]$  \\
  &              & Lee-Carter    & 2.54 $[1.51, 3.02]$ & 2.04 $[1.17, 2.55]$ & 2.03 $[1.53, 3.22]$ & 1.28 $[0.49, 4.16]$  & 2.07 $[1.75, 2.38]$  & 1.49 $[1.07, 1.73]$  \\
  &              & Li-Lee        & 2.02 $[0.62, 3.02]$ & 1.26 $[1.03, 1.86]$ & 2.03 $[1.53, 3.22]$ & 1.28 $[0.49, 4.16]$  & 2.06 $[1.75, 2.38]$  & 1.49 $[1.07, 1.73]$  \\
  &              & \textsc{mem5} & 7.85 $[6.21, 9.09]$ & 4.42 $[3.20, 5.85]$ & 7.37 $[5.51, 9.58]$ & 3.51 $[1.60, 17.08]$ & 5.59 $[4.24, 6.91]$  & 4.26 $[3.84, 5.31]$  \\
  &              & Oeppen        & 1.92 $[0.90, 3.39]$ & 1.74 $[1.16, 2.59]$ & 2.88 $[1.72, 5.01]$ & 2.59 $[0.81, 17.30]$ & 2.45 $[1.94, 10.05]$ & 2.03 $[1.37, 14.23]$ \\
  \midrule
  \parbox[t]{2mm}{\multirow{5}{*}{\rotatebox[origin=c]{90}{\textsc{males}}}} 
  & \textsc{mae} & \name{}       & 1                   & 1                   & 1                   & 1                   & 1                   & 1        \\
  &              & Hyndman-Ullah & 2.24 $[1.82, 2.44]$ & 2.47 $[2.09, 2.69]$ & 2.03 $[1.81, 2.35]$ & 0.71 $[0.65, 1.20]$ & 1.67                & 1.01 $[0.85, 1.11]$  \\
  &              & Lee-Carter    & 1.90 $[1.46, 2.36]$ & 2.43 $[1.88, 2.67]$ & 1.98 $[1.54, 2.31]$ & 0.69 $[0.62, 1.26]$ & 1.64 $[1.09, 2.34]$ & 0.97 $[0.69, 1.10]$  \\
  &              & Li-Lee        & 1.90 $[1.46, 2.36]$ & 2.43 $[1.88, 2.67]$ & 1.98 $[1.54, 2.31]$ & 0.69 $[0.62, 1.26]$ & 1.64 $[1.09, 2.34]$ & 0.97 $[0.69, 1.10]$  \\
  &              & \textsc{mem5} & 3.54 $[2.67, 3.92]$ & 3.71 $[3.08, 4.23]$ & 3.11 $[2.79, 3.59]$ & 1.12 $[0.98, 1.57]$ & 2.12 $[1.27, 3.76]$ & 1.59 $[1.09, 1.91]$  \\
  &              & Oeppen        & 1.25 $[1.06, 1.80]$ & 2.37 $[1.82, 2.83]$ & 1.54 $[1.20, 2.16]$ & 1.17 $[0.66, 1.69]$ & 1.78 $[1.07, 2.55]$ & 0.93 $[0.69, 1.10]$  \\
	  \cmidrule{2-9}                                                                                                                                                                                                
  & \textsc{mse} & \name{}       & 1                     & 1                     & 1                    & 1                   & 1                    & 1                 \\
  &              & Hyndman-Ullah & 5.90 $[4.18, 7.15]$   & 5.20 $[4.11, 7.28]$   & 4.01 $[3.29, 5.30]$  & 0.64 $[0.45, 1.85]$ & 2.95 $[1.18, 6.27]$  & 1.22 $[0.82, 1.45]$  \\
  &              & Lee-Carter    & 4.23 $[2.54, 6.61]$   & 5.00 $[3.35, 7.04]$   & 3.54 $[2.84, 5.13]$  & 0.61 $[0.43, 2.03]$ & 2.88 $[1.35, 5.69]$  & 1.10 $[0.55, 1.42]$  \\
  &              & Li-Lee        & 4.23 $[2.54, 6.61]$   & 5.00 $[3.35, 7.04]$   & 3.54 $[2.84, 5.13]$  & 0.61 $[0.43, 2.03]$ & 2.88 $[1.35, 5.69]$  & 1.10 $[0.55, 1.42]$  \\
  &              & \textsc{mem5} & 14.42 $[8.66, 17.81]$ & 10.93 $[8.03, 16.34]$ & 9.25 $[8.18, 12.72]$ & 1.40 $[0.93, 2.83]$ & 4.07 $[1.90, 13.27]$ & 2.69 $[1.25, 3.80]$  \\
  &              & Oeppen        & 2.65 $[1.60, 9.93]$   & 5.08 $[3.64, 7.59]$   & 3.36 $[1.78, 5.38]$  & 1.90 $[0.72, 3.24]$ & 3.51 $[1.31, 6.38]$  & 1.14 $[0.60, 1.60]$  \\
  \bottomrule
	\end{tabular}
	\caption{Out-of-sample relative performance. Median across rolling windows, stratified by country (\textsc{fra}, \textsc{gbr}, \textsc{ita}, \textsc{nld}, \textsc{nor}, \textsc{swe}). First and third quartiles in squared brackets.}
	\label{app:oos_full2}
\end{sidewaystable}